\documentclass[useAMS,usenatbib]{mnras}
\newcommand{\Sun}{\ensuremath{\odot}}

\usepackage{siunitx}
\usepackage{hyperref}
\usepackage{epsfig}
\usepackage{amsmath}
\usepackage{amssymb}
\usepackage{graphicx}
\usepackage{multirow}
\usepackage{arydshln}
\usepackage{url}
\usepackage[usenames,dvipsnames]{color}
\usepackage[normalem]{ulem}
\usepackage[T1]{fontenc}
\definecolor{gray}{gray}{0.7}
\usepackage[normalem]{ulem} 
\usepackage{verbatim}

\DeclareMathOperator{\erf}{erf}

\title[Lyman-$\alpha$ in the CGM]
{Lyman-$\alpha$ absorption beyond the disk of simulated spiral galaxies}
\author[B.~R\"ottgers et al.]
{Bernhard R\"ottgers$^{1}$,
Thorsten Naab$^{1}$\footnotemark,
Miha Cernetic$^{1}$,
Romeel Dav\'e$^{2,3,4}$, \newauthor
Guinevere Kauffmann$^{1}$,
Sanchayeeta Borthakur$^{5}$,
Horst Foidl$^{6}$
\\
$^1$Max-Planck Institut f\"ur Astrophysik, Karl-Schwarzschild-Str. 1,
85748 Garching, Germany \\
$^2$Institute for Astronomy, Royal Observatory, Univ. of Edinburgh, Edinburgh EH9 3HJ, UK\\
$^3$University of the Western Cape, Bellville, Cape Town 7535, South Africa\\
$^4$South African Astronomical Observatories, Observatory, Cape Town 7925, South Africa\\
$^5$School of Earth and Space Exploration, Arizona State University, 781 Terrace Mall, Tempe, AZ 85287, USA\\
$^6$Department of Astrophysics, University of Vienna, T\"urkenschanzstr. 17, 1180 Wien, Austria
}
\date{Accepted ???. Received ??? in original form ???}

\pagerange{\pageref{firstpage}--\pageref{lastpage}} \pubyear{2013}

\begin{document}
\label{firstpage}
\maketitle

\begin{abstract}
We present an analysis of the origin and properties of the circum-galactic medium (CGM) in a suite of 11 cosmological zoom simulations resembling present day spiral galaxies. On average the galaxies retain about 50\% of the cosmic fraction in baryons, almost equally divided into disc (interstellar medium) gas, cool CGM gas and warm-hot CGM gas. At radii smaller than 50~kpc the CGM is dominated by recycled warm-hot gas injected from the central galaxy, while at larger radii it is dominated by cool gas accreted onto the halo. The recycled gas typically accounts for one-third of the CGM mass. We introduce the novel publicly available analysis tool \textsc{pygad} to compute ion abundances and mock absorption spectra. For Lyman-${\alpha}$ absorption we find good agreement of the simulated equivalent width (EW) distribution and observations out to large radii. Disc galaxies with quiescent assembly histories show significantly more absorption along the disc major axis. By comparing the EW and HI column densities we find that CGM Lyman-${\alpha}$ absorbers are best represented by an effective line-width $b\approx 50 - 70$~km~s$^{-1}$ that increases mildly with halo mass, larger than typically assumed.   
\end{abstract}


\footnotetext{E-mail: naab@mpa-garching.mpg.de}


%
%


\section{Introduction}
\label{intro}
Star-forming galaxies mostly grow by accreting gas from their surrounding medium, fuelling star formation~\citep{2005MNRAS.363....2K,2009Natur.457..451D}. This accretion can include relatively pristine gas that has comes directly from the intergalactic medium (IGM), or gas that was previously ejected from a galaxy, known as wind recycling~\citep{2010MNRAS.406.2325O}, including potentially significant intergalactic transfer~\citep{2017MNRAS.470.4698A}. The circum-galactic medium (CGM) thus serves as a  reservoir that holds signatures of accretion, outflows, and wind recycling, as well as holding the majority of baryons within halos~\citep{2014ApJ...786...54P,2017ARA&A..55..389T}.

At present, the CGM is best probed using absorption line techniques. From HI and metal absorption studies around low-redshift galaxies, the CGM is found to be highly multi-phase~\citep[e.g.][]{2014ApJ...792....8W,2019MNRAS.488.1248H,2019arXiv191001123H} in a manner that depends on the host galaxy properties~\citep[e.g.][]{2011Sci...334..948T,2012ApJ...758L..41T}. It is likely to be dynamic, although it appears that absorber relative velocities rarely exceed the galaxy escape velocity~\citep{2013ApJ...777...59T}.  These complex processes can be difficult to disentangle using only the one-dimensional spectral probes along a limited number of sightlines.  For this reason, gas dynamical simulations are now commonly employed to help relate the observable absorption line properties with the physical and dynamical state of CGM gas~\citep{2013MNRAS.432...89F,2014MNRAS.444.1260F}.  

Modelling the CGM represents a substantial computational challenge. To begin with, it is important to have a model that is able to produce realistic galaxy properties \citep{2013MNRAS.432...89F,2016MNRAS.462.3751K,2017MNRAS.464.2796G,2018MNRAS.477..450N,2018MNRAS.481..835O,2019arXiv191001123H}.  Only in recent years have groups succeeded in simulating realistic spirals, owing primarily to improvements in understanding of the impact of star formation feedback (see \citealp{2015ARA&A..53...51S,2017ARA&A..55...59N} for reviews), and numerical resolution afforded by advancing computing power.  On a cosmological scale such models nonetheless make a wide range of predictions for CGM gas \citep{2016MNRAS.459.1745F,2016MNRAS.462.3751K,2018MNRAS.477..450N}, owing to variations in feedback physics that, while constrained to reproduce the stellar component, can deposit widely varying amounts of mass, metals, and energy into the CGM. It is also not clear that current CGM simulations have achieved convergence \citep{2019MNRAS.486.4686K}, even as clever techniques have been employed to enhance resolution specifically in the CGM~\citep{2019MNRAS.482L..85V,2019ApJ...873..129P, hummels_2019_apj}. The CGM is thus a crucial test bed for modern galaxy formation models, and developing improved models for examining the role of the CGM in galaxy evolution remains an important goal.

In this paper we examine the CGM physical conditions and associated absorption line properties in a suite of eleven zoom-in simulations of star-forming galaxies spanning from dwarfs to super-$L^\star$.  We employ a Smoothed Particle Hydrodynamics (SPH)-based code that explicitly tracks the cold and hot ISM phases separately \citep{2013MNRAS.434.3142A}. By tracking particles, we are able to quantify the origin of CGM gas, both pristine and enriched, in these simulations. We examine radial trends of absorption strength in various ions. Along the way, we introduce a new publicly-available code to generate absorption line spectra called \citep[{\sc pygad},][]{2018ascl.soft11014R} \footnote{\textsc{pygad} is available at: \mbox{\url{https://bitbucket.org/broett/pygad}}. The legacy version used in this paper is in Python 2. A Python 3 version is available and will be maintained in the future.}, which doubles as a general-purpose SPH simulation visualisation and analysis tool.  

This paper is organised as follows: in section \ref{simulations} we describe and briefly review the simulations. The bulk properties of the ISM and CGM are presented in section \ref{bulkproperties}.  In section 4 we describe our spectral line generation code, and in section 5 we use this to study CGM absorption properties. We discuss feedback and resolution dependence in section 6, and summarise in section 7.


\section{Simulations}
\label{simulations}

\subsection{Code Description}
\label{sec:code_description}
The simulations are performed with the multi-phase cosmological
smoothed particle hydrodynamics implementation of \textit{GADGET}
\citep{2005MNRAS.364.1105S} presented in
\citet{2013MNRAS.434.3142A}. It is a modified version of the code used
by \citet{2006MNRAS.371.1125S,2005MNRAS.364..552S} and is built on the
TreeSPH code \textit{GADGET-3}. The code includes models for a
multi-phase gas treatment, metal line cooling, star formation, stellar
metal production and feedback, and a spatially homogeneous,
redshift-dependent UV-background \citep{2001cghr.confE..64H}. With
this implementation it is possible to successfully produce 
realistic present day properties, { like stellar sizes, kinematic stellar disc mass fractions, and HI sizes} and evolution histories of the spiral galaxy population \citep{2013MNRAS.434.3142A,2014MNRAS.441.3679A,2014MNRAS.443.2092U,2014MNRAS.441.2159W}.
Here we briefly describe the unchanged features of the code and in more detail the added improvements with respect to energy conservation in all feedback events,
and the decoupling of the AGB feedback from the feedback of Type Ia supernovae (SN). 
For further details we refer the reader to \cite{2013MNRAS.434.3142A} and \cite{2006MNRAS.371.1125S,2005MNRAS.364..552S}.

\subsection{Multi-phase Gas Model}
\label{sec:phases}

From observations it is known that the interstellar gas
has a complex structure with different phases -- a hot volume-filling phase and a dense cold gas phase -- both of which should be represented in galaxy formation simulations (see \citealp{2017ARA&A..55...59N,2017ARA&A..55..389T}). 
To model this, we treat the gas as a multi-phase medium with
many co-existing phases (see \citealp{2003MNRAS.345..561M}).
We let two SPH particles $i$ and $j$ decouple into separate phases, if the
following two conditions apply \citep{2013MNRAS.434.3142A}: 
\begin{eqnarray}
\max \left( \frac{A_i}{A_j}, \frac{A_j}{A_i} \right) &> 50 \label{eq:decoupling_A}, \;\;\;\; 
-\mu_{ij} &< c_{ij} \label{eq:decoupling_vel}.
\end{eqnarray}
Here $A_{i,j}$ are the entropic functions of the particles
(\citealp{2002MNRAS.333..649S}), \mbox{$\mu_{ij} := (\vec v_i - \vec
  v_j) \cdot \frac{\vec r_i - \vec r_j}{|\vec r_i - \vec r_j|}$} is
the relative velocity of the particles along their vector of
separation, and $c_{ij}$ is the pair-averaged sound-speed. Two
SPH-particles decouple if their entropy (actually their entropic
functions\footnote{The entropic function  
$A$ is defined by $P =: A(s) \rho^\gamma$, where $P$ is the pressure,
  $s$ is the specific entropy, $\rho$ is the density, and $\gamma$ is
  the adiabatic index.}) are very different unless they approach
faster than with the local sound speed. The velocity restriction
(Eq. \ref{eq:decoupling_vel}) is required to capture shocks properly
\citep{2003MNRAS.345..561M}. This multi-phase treatment results in a continuum of phases from cold to hot and the results are not
very sensitive to the exact ratio in
Eq.~(\ref{eq:decoupling_A}). { This model is aimed at preventing overcooling, i.e. artificially short cooling times \citep[see e.g.][for a review]{2017ARA&A..55...59N} and allows for the simultaneous representation of, and energy injection into, a hot and a cold phase on the resolution scale \citep[see][for a detailed discussion]{2006MNRAS.371.1125S}. Such multi-phase ISM structure naturally arises in much higher resolution simulations of the supernova driven multi-phase ISM \citep[see e.g.][]{2015MNRAS.454..238W}}. For further details on the multi-phase model see \citet{2003MNRAS.345..561M,2013MNRAS.434.3142A,2014MNRAS.441.3679A}.

\subsection{Star Formation}

Gas particles are converted into star particles stochastically such
that on average they have a star formation rate of 
\begin{align}
\mathrm{SFR} = \eta \, \frac{\rho_\mathrm{gas}}{t_\mathrm{dyn}} \quad \Big(\propto \rho_\mathrm{gas}^{1.5}\Big). 
\end{align}
Here $\rho_\mathrm{gas}$ is the gas density and $t_\mathrm{dyn} \equiv
\left(4 \pi G \rho_\mathrm{gas} \right)^{-1/2}$  is the local
dynamical time for the gas particle. This results in a
Kennicutt-Schmidt like relation
\citep{1959ApJ...129..243S,1998ApJ...498..541K} locally and the star
formation efficiency $\eta$ is set to a typical value of 0.04
\citep{2013MNRAS.434.3142A}.   

Gas is only allowed to form stars if the density fulfils the following
conditions \citep{2013MNRAS.434.3142A}:
\begin{align}
n &> n_\mathrm{th} \simeq 3 \, \mathrm{cm}^{-3}\simeq 10^{-1}M_\Sun/{\rm pc}^3, \;\;\;\;
\frac{\rho}{\langle\rho\rangle} &> 2000,
\end{align}
where $\langle\rho\rangle$ is the cosmic mean {}{baryon} density
and $n$ is the number density of the gas.

Each stellar particle spawned from the gas represents a stellar population with a
\cite{2001MNRAS.322..231K} initial mass function in the range of $0.1 \, \mathrm{M_\Sun}$ and $100\, \mathrm{M_\Sun}$. 

\subsection{Metal Enrichment}


\subsubsection{Metal Production}
\label{sec:metal_prod}

We follow enrichment and cooling for H, He, and nine metals: carbon, nitrogen, oxygen, neon, magnesium, silicon, sulphur, calcium and iron, as they are
produced by type Ia, II supernovae as well as AGB stars. The SNe II yields are metal-dependent following
\cite{2004ApJ...608..405C}, but with the iron yield halved as proposed
by \cite{1998A&A...334..505P}.  
The element production by SNe Ia follows the W7 model of
\cite{1999ApJS..125..439I}, with a declining rate by the inverse of
the stellar population age sampled in time steps of
50~Myr. We account for the mass recycling by AGB stars, with the same
time sampling as for SNe Ia, by using the metal yields of
\cite{2010MNRAS.403.1413K}. In contrast to \cite{2013MNRAS.434.3142A}
we do not add the mass return of AGB stars to that of the supernovae
but assume an outflow velocity of $30\,\mathrm{km/s}$. This reduces the
energy input by AGB winds, but does not change the results significantly (e.g.~compare Fig.~\ref{fig:smhm}). The galaxies become smaller which is in better agreement with observations.

For the feedback we distinguish two different gas phases (again
following \citealp{2013MNRAS.434.3142A}), with
a cold phase defined by \mbox{$T < 8\times10^{4}\,\mathrm{K}$} and \mbox{$n > 4\times10^{-5}\,\mathrm{cm}^{-3}$}), and the remaining gas in the hot phase.  Note these phases are distinct from the range of phases used for hydrodynamics (cf.\ Eq.~\ref{eq:decoupling_A}).
Metals and energy/momentum from AGBs and SNe are returned
separately into these two gas phases. In our implementation 50\%
of the metals created in a supernova are deposited into the
neighbouring hot phase, the other 50\% are deposited in the cold phase.
We proceed in the same way for momentum and energy injection (see also
\S~\ref{sec:feedback}). The motivation is that in the neighbourhood (i.e. within a kernel, that
contains about $2.4 \times 10^7 \mathrm{M_\Sun}$, spread over at least $\sim 200\,\mathrm{cpc/h}$ of a star, there is always unresolved hot and cold dense gas and any feedback event will
typically deposit mass and energy into both phases. 
The exact ratios are unknown in nature. \cite{2013MNRAS.434.3142A} found that the simple choice of 50:50
yields reasonable galaxies in terms of the global measures such as the stellar
mass-halo mass relation.

\subsubsection{Metal Line Cooling}

As shown by \cite{2009MNRAS.393...99W}, cooling rates by metal lines with the correct 
element abundances are important for a correct treatment in galaxy formation simulations.
We implement their rates for optically thin gas in ionisation equilibrium.
This includes a redshift dependent UV background of \cite{2001cghr.confE..64H} making the cooling rates dependent on redshift, density, temperature, and chemical composition.

\subsubsection{Metal Diffusion}

We follow metal diffusion due to turbulent gas mixing with an SPH formulation of the diffusion equation
\mbox{($\frac{\mathrm d c}{\mathrm d t} = \frac{1}{\rho} \vec\nabla \cdot (D \vec\nabla c)$}
for some concentration $c$ and a diffusion coefficient $D$).
We do not simply use a discretized SPH formulation, but follow
\cite{2009MNRAS.392.1381G}, who argued for a version integrated over the small
discrete time step $\Delta t$ of the simulation. To conserve (metal) mass we use the same equation for the metal mass
$\mu_i = c_i m_i$ of particle~$i$ as \cite{2013MNRAS.434.3142A}:
\begin{align}
    \Delta \mu_i &= \sum_j \mu_{i \rightarrow j} \nonumber \\
    &= \sum_j \left( \frac{1}{2} m_i \left( 1 - e^{A \Delta t} \right) \frac{1}{A}
    K_{ij} (c_i - c_j) \right),
\end{align}
where the sum goes over all SPH neighbours~$j$ of particle~$i$, $m_i$ is the
$i$-th particle's mass, \mbox{$A \equiv \sum_j K_{ij}$}, and
\begin{align}
    K_{ij} = \frac{m_j}{\rho_i \rho_j} \frac{4 D_i D_j}{D_i + D_j} \frac{\vec
    r_{ij} \cdot \vec\nabla_i W_{ij}}{r_{ij}^2}.
\end{align}
Here $\rho_{i,j}$ are the respective particle densities, $W_{ij}$ is the kernel of particle~$i$ at the position of particle~$j$,
and $\vec r_{ij}$ is the separation vector of the two particles.

Finally, we choose the diffusion coefficients $D_i$ as proposed by \cite{2010MNRAS.407.1581S}, 
which does not introduce diffusion in purely rotational or compressive flows due to the use of the 
trace-free tensor $S_{kl}$ (see their paper for details):
\begin{align}
    D_i = 0.05 \rho_i |S_{kl}| h_i^2,
\end{align}
where $h_i$ is the smoothing length of particle~$i$.
\cite{2013MNRAS.434.3142A} found that this coefficient also yields better
agreement with observational data, such as the mass-metallicity relation, when
applied in cosmological galaxy formation simulations than the one proposed by \cite{2009MNRAS.392.1381G}.

\subsection{Stellar Feedback}

We consider thermal and kinetic feedback from SNe II, SNe Ia, and winds from AGB stars.
In this section we describe how the feedback is modelled, and the 
modifications to \cite{2013MNRAS.434.3142A}. We also briefly describe
the additional approximation to radiative feedback from O and B stars.

\subsubsection{Thermal and Kinetic Feedback}
\label{sec:feedback}

We assume that each SN injects $10^{51}\,\mathrm{erg}$ into the surrounding ISM.
Each of the 10 nearest hot and cold neighbours receive the momentum corresponding to an outflow velocity of $3000~\mathrm{km}\,\mathrm{s}^{-1}$ pointing radially away from the
stellar population particle. The transferred momentum changes the
kinetic energy of the receiving gas particle. Assuming inelastic
collisions, the remaining energy is added as thermal energy to ensure
energy conservation. The thermal energy is not added to
particles of the hot phase, but put into a reservoir for cold phase
particles (using the definition in section~\ref{sec:phases}). 
Once the reservoir energy is sufficient to heat (`promote') the
particle to the hot phase, it is released (i.e.\ added to the thermal
energy). For a more detailed description we refer to
\cite{2013MNRAS.434.3142A}. 

\subsubsection{Radiation pressure}

Based on \cite{2011MNRAS.417..950H} the stellar populations deposit a momentum of
\begin{align}
\dot p_\mathrm{rp} = (1 + \tau_\mathrm{IR}) \frac{L_\mathrm{UV}}{c},
\end{align}
to the 10 nearest neighbours as a continuous force acting over the first 30~Myr in the life cycle of a stellar population particle.
Here $L_\mathrm{UV}$ is the UV luminosity, $c$ is the speed of light, and $\tau_\mathrm{IR}$ is the infrared optical depth. 
The latter is modelled by
\begin{align}
\tau_\mathrm{IR} = \tau_0 \cdot \max \left( 1, \frac{\rho h}{\rho_\mathrm{sf} h_\mathrm{sf}} \right) \cdot \left( \frac{Z}{Z_\Sun} \right) \cdot \min\left( \left( \frac{\sigma}{\sigma_0} \right)^3, 4 \right),
\end{align}
i.e. scaling with the local metallicity $Z$, the particle's surface
density approximated by its density $\rho$ times its smoothing length
$h$ (limited by the star forming surface density threshold
$\rho_\mathrm{sf} h_\mathrm{sf}$), and the surrounding
velocity dispersion $\sigma$ cubed.
The parameters used are $\tau_0 = 25$ and $\sigma_0 =
40\,\mathrm{km}\,\mathrm{s}^{-1}$. Additionally,
the factor of $\left(\sigma / \sigma_0 \right)^3$ is limited to $4$,
in order to avoid overly strong forces.
This relatively strong radiation pressure feedback helps dispersing dense star forming
regions and results in formation histories and final stellar masses in
agreement with observations (cf.\ also Fig.~1 of \citealp{2013MNRAS.434.3142A}).

\subsection{The Simulation Sample}

\begin{figure}
\centering
\includegraphics[width = 0.49\textwidth]{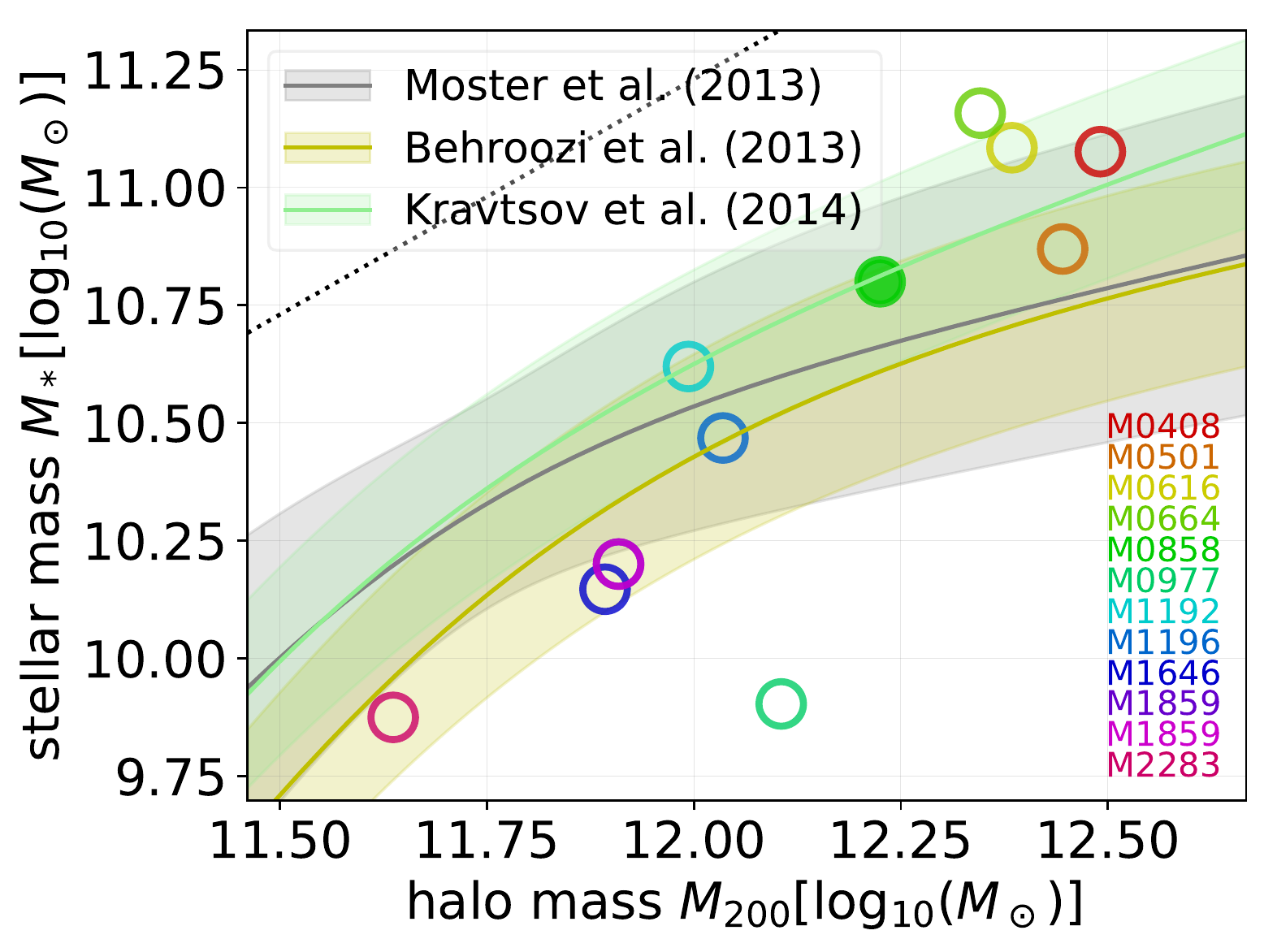}
\caption{The stellar mass--halo mass relation for the simulated halos at redshift zero. The example halo discussed in detail (M0858) is highlighted as a solid (green) circle. 
	For reference, the abundance matching results from
        \citealp{2014arXiv1401.7329K, 2013ApJ...770...57B,
          2013MNRAS.428.3121M} are shown in differently coloured bands.
	With the exception of M0977, which is undergoing a major merger
	at $z=0$ (see text for more detail), all central galaxies agree with
	abundance matching constrains.
	At higher halo masses the simulated stellar masses are high, possibly
	due to missing black hole feedback.
	The dashed black line indicates the cosmic baryon fraction of the given halo.
}
\label{fig:smhm}
\end{figure}

\begin{table*}
\centering
\begin{tabular}{l|ccccccccc|l}
halo & $R_{200}$ & $M_{200}$ & $M_*$ & $f_\mathrm{gas}$ & Z$_\mathrm{gas}$ & $M_\mathrm{HI}$ & $R_\mathrm{HI}$ & $M_\mathrm{CGM}$ & $f_\mathrm{CGM}^\mathrm{accreted}$ &  late merger \\
ID & [kpc] & [$\mathrm{10^{12}\ M_\Sun}$] & [$\mathrm{10^{11}\ M_\Sun}$] & [$\%$] &  [$Z_\Sun$] & [$\mathrm{10^{10}\ M_\Sun}$] & [kpc] & [$\mathrm{10^{11}\ M_\Sun}$] & [$\%$] &  history \\
\hline
M0408   & 295 & $3.10$ & 1.19 & 21.6 & 2.36 & 2.12 & 25 & 1.56 & 72.8 & smaller, $z\approx0.3$ \\
M0501   & 285 & $2.79$ & 0.74 & 16.7 & 1.94 & 1.01 & 62 & 1.07 & 67.2 & large, $z\approx0.9$ \\
M0616   & 272 & $2.42$ & 1.22 & 20.7 & 3.02 & 1.78 & 17 & 1.31 & 70.0 & smaller, $z\approx0.05$ \\
(M0664) & 264 & $2.22$ & 1.44 & 7.2  & 2.72 & 0.45 & 10 & 1.08 & 84.2 & ongoing major \\
M0858   & 241 & $1.68$ & 0.63 & 22.4 & 2.16 & 1.15 & 17 & 0.70 & 65.3 & small fly-by, $z\approx0.1$ \\
(M0977) & 220 & $1.27$ & 0.08 & 55.9 & 0.50 & 0.69 & 45 & 0.82 & 86.2 & ongoing major \\
M1192   & 201 & $0.99$ & 0.42 & 22.2 & 2.01 & 0.81 & 20 & 0.22 & 70.9 & smaller, $z\approx0.2$ \\
M1196   & 208 & $1.08$ & 0.29 & 26.9 & 1.43 & 0.76 & 25 & 0.30 & 74.9 & large, $z\approx0.4$ \\
(M1646) & 186 & $0.78$ & 0.14 & 43.8 & 1.04 & 0.75 & 20 & 0.28 & 77.3 & small fly-by, $z\approx0.1$ \\
M1859   & 189 & $0.81$ & 0.16 & 36.4 & 1.05 & 0.64 & 25 & 0.15 & 73.2 & small, $z\gtrsim0.5$ \\
(M2283) & 153 & $0.43$ & 0.07 & 32.8 & 0.81 & 0.23 & 11 & 0.16 & 64.2 & ongoing major \\
\end{tabular}
\caption{
	Overview of the simulated halos at redshift $z=0$ used for
    this study. The naming (IDs) is taken from     \citet{2010ApJ...725.2312O,2012ApJ...744...63O} and is identical to \citet{2013MNRAS.434.3142A}.
    $R_{200}$ is the radius around the galaxy where the mean spherical overdensity drops below $200\,\rho_\mathrm{crit}$ and $M_{200}$ is the mass within. $M_*$ is the stellar mass within $10\%$ of $R_{200}$. 
    Gas fractions ($f_\mathrm{gas}$), metallicity ($Z_{\mathrm{gas}}$), and hydrogen mass ($M_\mathrm{HI}$) are also taken within that radius. 
    {}{$R_\mathrm{HI}$ is the size of the H\textsc{i} disc, defined as the radius at which surface density of H\textsc{i} falls below $1\,\mathrm{M_\Sun} \mathrm{pc}^{-2}$.}
    The mass of the CGM ($M_\mathrm{CGM}$) is measured within $R_{200}$ (see text) and $f^{\mathrm{accreted}}_{\mathrm{CGM}}$ is the CGM mass fraction which is accreted onto the halo from the outside. Galaxies with ongoing (major) galaxy mergers are in parentheses and are excluded from the analysis.}
\label{tab:haloprops}
\end{table*}

We simulated a suite of 11 zoomed-in
cosmological simulations in a box of $(100~\mathrm{cMpc})^3$, from the parent
dark matter-only simulation described in
\cite{2010ApJ...725.2312O}. It was run with a flat $\Lambda$CDM
cosmology with \mbox{$H_0 =
  72\,\mathrm{km}\,\mathrm{s}^{-1}\,\mathrm{Mpc}^{-1}$},
\mbox{$\Omega_\Lambda=0.74$}, \mbox{$\Omega_\mathrm{matter}=0.26$},
\mbox{$\Omega_\mathrm{bayron}=0.044$}, \mbox{$\sigma_8 = 0.77$, and
  $n_s = 0.95$}. 

For the zoom simulations we use the same initial conditions (ICs)
and naming convention as \cite{2010ApJ...725.2312O} and
\cite{2013MNRAS.434.3142A}. We restrict ourselves to ICs of halos
with present day virial masses in the range of \mbox{$M_{200} \in
[4\times10^{11}\,\mathrm{M_\Sun}, \  3\times10^{12}\,\mathrm{M_\Sun}]$}.  This
corresponds to stellar masses in the range from
\mbox{$7.5\times10^{9}\,\mathrm{M_\Sun}$} to \mbox{$1.2\times10^{11}\,\mathrm{M_\Sun}$}
(cf.\ Fig.~\ref{fig:smhm}).

With the minor changes to the original code of
\cite{2013MNRAS.434.3142A}, the key properties of the galaxies do not
change significantly. As an example we show the stellar mass-halo mass
relation at $z=0$ (see Fig.~\ref{fig:smhm}).
Except for halo M0977 all central galaxies fall well into the region for stellar
masses expected from abundance matching results.
M0977 is in a late major merger stage
and its stellar mass is about to double from
$8\times10^{9}\,\mathrm{M_\Sun}$ to $1.4\times10^{10}\,\mathrm{M_\Sun}$.

The galaxies in the most massive halos ($> 2\times10^{12}\,\mathrm{M_\Sun}$) of our
ensemble have a tendency towards higher stellar masses, which might be attributed
to the absence of super-massive black holes (SMBH) and the feedback of active
galactic nuclei (AGN) (see e.g. \citealp{2015ARA&A..53...51S,2017ARA&A..55...59N}).

\begin{figure*}
\centering
\includegraphics[width = \textwidth]{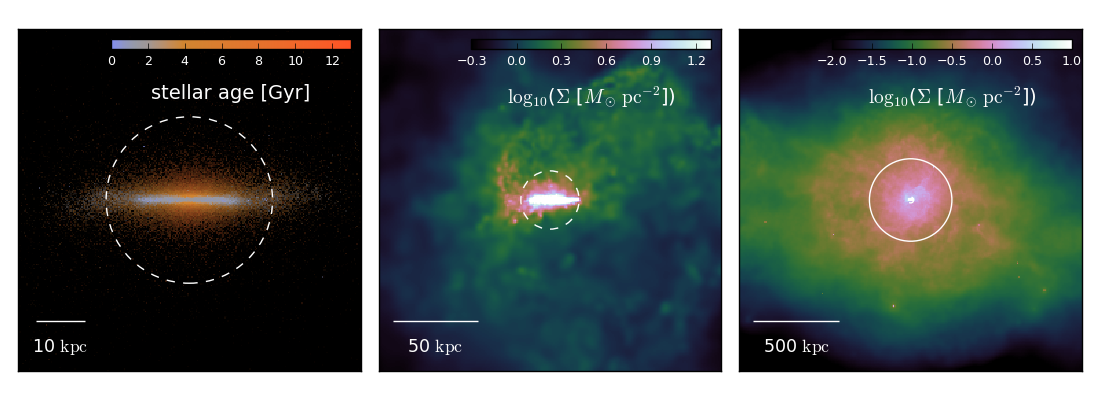}
\caption{
    Morphological impression of galaxy M0858. In the left panel
    we show (edge-on) the stellar V-band luminosities (determining the pixel
    brightness / the value in HSV colour space), colour coded by the mean
    (V-band weighted) stellar ages (determining the pixel hue).
    The galaxy has a thin and young stellar disk.
    The middle panel displays the gas column density distribution in and around
    the galaxy. In the right panel we show the gas distribution beyond the
    galaxies' virial radius ($R_{200}$), which is indicated by the solid circle.
    The dashed circles in the other panels is at the H\textsc{i} disc
    size, $R_\mathrm{HI}$. The respective spatial scales are given in the lower
    left corner of the panels.
}
\label{fig:overview_M0858}
\end{figure*}

{The galaxies have large dynamical stellar disc mass fractions as listed in Table~\ref{tab:haloprops}, ranging from about one third to more than $80\%$. 
Among these values, }{}Table~\ref{tab:haloprops} lists {other}{} key features of the galaxies at redshift zero such as the H\textsc{i} disc size, which is of interest for us as we investigate Lyman-$\alpha$
absorption. {}{For details on the H\textsc{i} disc sizes see Sec.~\ref{sec:ionisation_fracs}.} The gas phase metallicities of ongoing major mergers can become very low due to merger driven metal poor inflow \citep[see e.g.][]{2009ApJ...695..259P}. These systems are excluded from the analysis. The calculation of these quantities is discussed in sections~\ref{sec:postprocess} and \ref{sec:ionisation_fracs}. 
As an example of our simulations we show the structure of halo M0858 in Fig.~\ref{fig:overview_M0858}{, the halo which we will also discuss in more detail as an example in the following.}{We will discuss and use this halo as an example in this work.}

\begin{table}
\centering
\begin{tabular}{c|cc|cc}
\multirow{2}{*}{level} & \multicolumn{2}{c|}{softening length [cpc/$h_0$]} & \multicolumn{2}{c}{particle masses [$M_\Sun$]} \\
  & baryons & dark matter & baryons & dark matter \\
\hline
2x & 400 & 900 & $5.9 \times 10^{6}$ & $2.9 \times 10^{7}$ \\
4x & 200 & 450 & $7.4 \times 10^{5}$ & $3.6 \times 10^{6}$ \\
8x & 100 & 225 & $9.2 \times 10^{4}$ & $4.5 \times 10^{5}$ \\
\end{tabular}
\caption{Overview of the different resolutions levels in the zoom region. The baryonic masses are initial masses, 
	since gas particles can later vary by a factor $\sim2$ as they accumulate mass from stellar
        feedback and metal diffusion.
}
\label{tab:resolutions}
\end{table}

We performed simulation at three different spatial
resolutions which are referred to as `2x', `4x', and `8x' (see
Table~\ref{tab:resolutions}). If not stated otherwise, we use the
intermediate resolution `4x'.
For `4x' the (initial) gas particle mass is
$7.4 \times 10^{5}\, \mathrm{M_\Sun}$ and the gravitational softening length is
$200\,\mathrm{cpc}/h_0$.

\subsection{Simulation Analysis}
\label{sec:postprocess}

All data analysis presented in this paper is performed with the publicly available, modular analysis tool \textsc{pygad}.

We identify the main halos and their largest galaxies of the high-resolution part of the zoom
simulations with a friends-of-friends (FoF) algorithm. The simulation snapshots are then centred on the main galaxies of these halos, using a shrinking sphere method
\citep{2003MNRAS.338...14P} on the stars.
We calculate the spherical overdensity mass $M_{200}$ and radius $R_{200}$
of the dark matter halos with respect to this centre.
All baryons within $10\%$ of $R_{200}$ are defined as galaxies (following \citealp{2010ApJ...725.2312O}). 

For rotating the systems to the edge-on projection we use the eigenvectors of the reduced moment-of-inertia tensor \citep{1983MNRAS.202.1159G,2005ApJ...627..647B} of the
galaxies, which are typically dominated by the stars as the mean gas fraction is close to $20\%$ (cf.\ Table~\ref{tab:haloprops}). The $z$-axis is then aligned with the minor axis and the $x$-axis along the major
axis of the galaxies. 

%

\section{ISM and CGM Bulk Properties}
\label{bulkproperties}
\subsection{Mass Budget}

The sum of observed mass components within galaxy halos has long been known to fall short of the expected cosmic baryon fraction, known as the missing halo baryon problem.  It has been suggested that the shortfall lies in multi-phase CGM gas, which has been difficult to quantify.  \cite{2014ApJ...792....8W} used COS-Halos and other data to estimate the cool and warm-hot CGM components within $\sim 10^{12}\,\mathrm{M_\Sun}$ halos to be 30-50\% and 5-40\%, respectively, with about 20\% in the ISM.  In contrast, \citet{2016ApJ...830...87S} determined a data-concordant universal CGM density profile and estimated 5-10\% of the baryons in the cool CGM phase.  

\begin{figure}
\centering
\includegraphics[width = 0.49\textwidth]{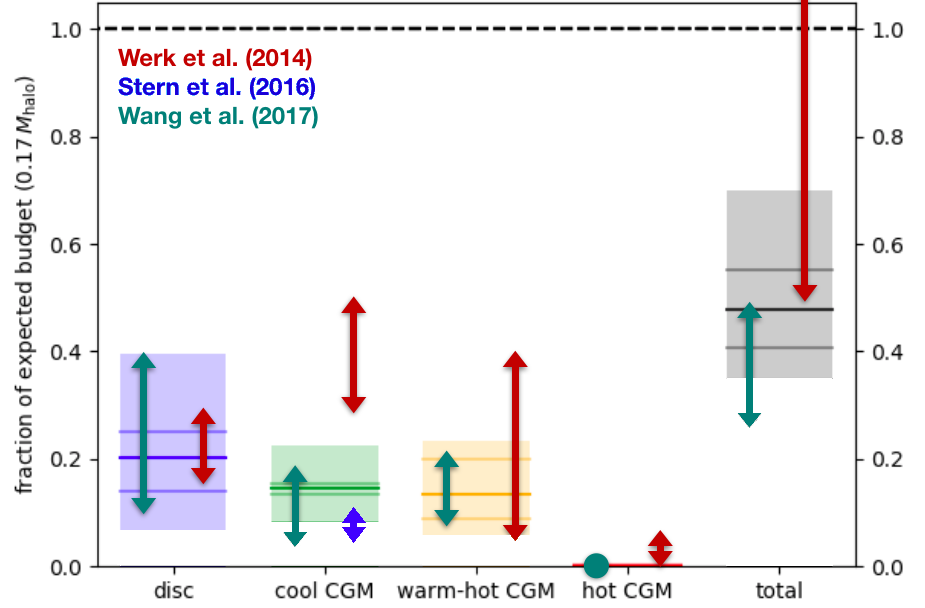}
\caption{
	The mass budget of the halos and the CGM in the simulations in a similar fashion as done by 
	\protect\cite{2014ApJ...792....8W} (indicated with red arrows here).
	Note that there are different measurement from \protect\cite{2016ApJ...830...87S} and 
	\protect\cite{2017MNRAS.466.4858W} for cool gas in 
	better agreement with our results.
	`disc' is all stars and the ISM gas in the halos; `cool CGM' is all remaining gas with $T < 10^5\,\mathrm{K}$;  
	`hot-warm CGM' gas has temperatures of $10^5\,\mathrm{K} \lesssim T \lesssim 10^7\,\mathrm{K}$; and `hot CGM' finally is gas with $T > 10^7\,\mathrm{K}$. The shaded region indicates the range of fractions found in the simulations. The median is marked with a thick line and the 25- and 75-percentiles are shown by the thin lines.}\label{fig:mass_budget}
\end{figure}

We compare our simulations with these estimates. We define gas within 15\% of $R_{200}$, $T<2\times10^4\,\mathrm{K}$ and
$n_H>10^{-2}\,\mathrm{cm}^{-3}$ as ISM gas (note that this might include the ISM of close satellites). The estimate for `disc' then includes ISM gas and the stellar component of the galaxy. Remaining gas within $R_{200}$ is considered CGM gas, divided into cool ($T < 10^5\,\mathrm{K}$), warm-hot 
($10^5\,\mathrm{K} < T < 10^7\,\mathrm{K}$), and hot ($10^7\,\mathrm{K} < T$) components. 
The results for our simulation suite is shown in  Fig.~\ref{fig:mass_budget} compared to observational estimates from \cite{2014ApJ...792....8W} and \citet{2016ApJ...830...87S}. The lightly shaded region shows the full range among the 11 zooms, with the median indicated by the thick horizontal line and the 25--75\% region shown by the thin lines.  

We find 20\% in the ISM, in agreement with observational determinations. The fraction in the cool and warm-hot CGM are slightly lower but similar at 15--25\% each. This results in a total baryon fraction of about 50\% of the expected baryon fraction, with the rest having been expelled outside the halo.  Our results are generally in agreement with observations, although the cool CGM estimate of \cite{2014ApJ...792....8W} is clearly higher; our results are in somewhat better agreement with \citet{2016ApJ...830...87S}.  For comparison, we also show the NIHAO simulation results from \citet{2017MNRAS.466.4858W}, which are generally similar to ours.

In summary it appears that modern galaxy formation models that reproduce the stellar-to-halo mass relation result in significant expulsion of gas from halos, and result in baryons being roughly equally divided between the ISM, the cool CGM, and the warm-hot CGM.  This is broadly concordant with observations, though the uncertainties remain substantial.

\subsection{Particle Tracing and CGM Gas History}

An interesting question is, how was the CGM assembled?  Part of it comes from accretion onto the halo, while some of it may arise from material ejected from the ISM.  To examine this, we take a prototypical galaxy from our sample, M0858, and trace the individual gas particles in time to examine the relative contribution from these two channels.


\begin{figure}
\centering
\includegraphics[width = 0.5\textwidth]{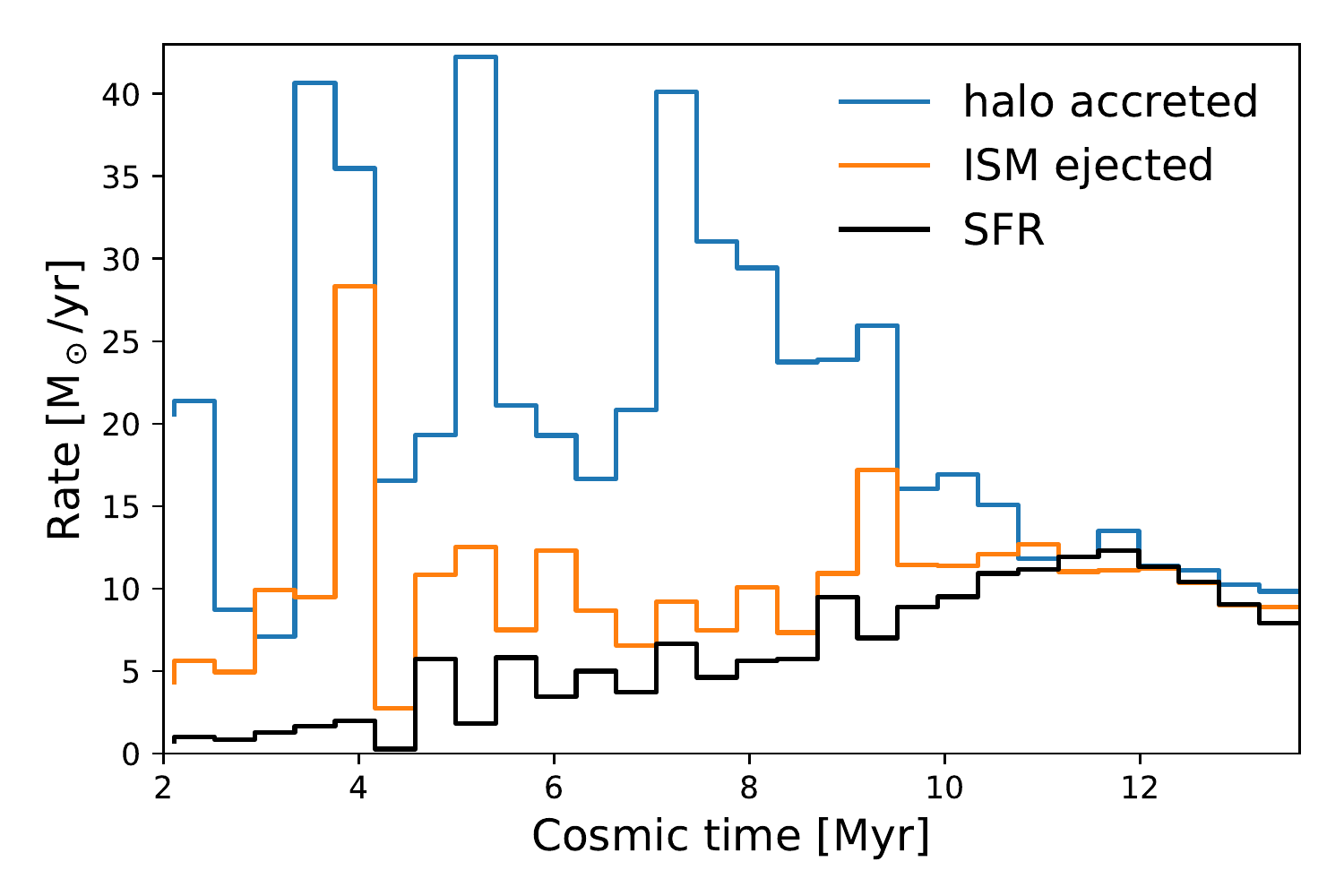}
\caption{
	The {}{instantaneous} accretion and ejection histories of the CGM gas of M0858 over cosmic time.
	{Orange}{Blue} is gas crossing $R_{200}$ on its way into the halo and becoming part of the CGM.
	 {The blue}{Orange} {histogram shows}{is} gas ejected from the ISM (see text for details) into the CGM. {}{The star formation rate in the galaxy (within 10\% of R200) is shown in black. At late cosmic times the star formation and ejection rate becomes comparable}
}
\label{fig:M0858_ism_halo_ejection_hist}
\end{figure}

\begin{figure*}
\centering
\includegraphics[width = 0.49\textwidth]{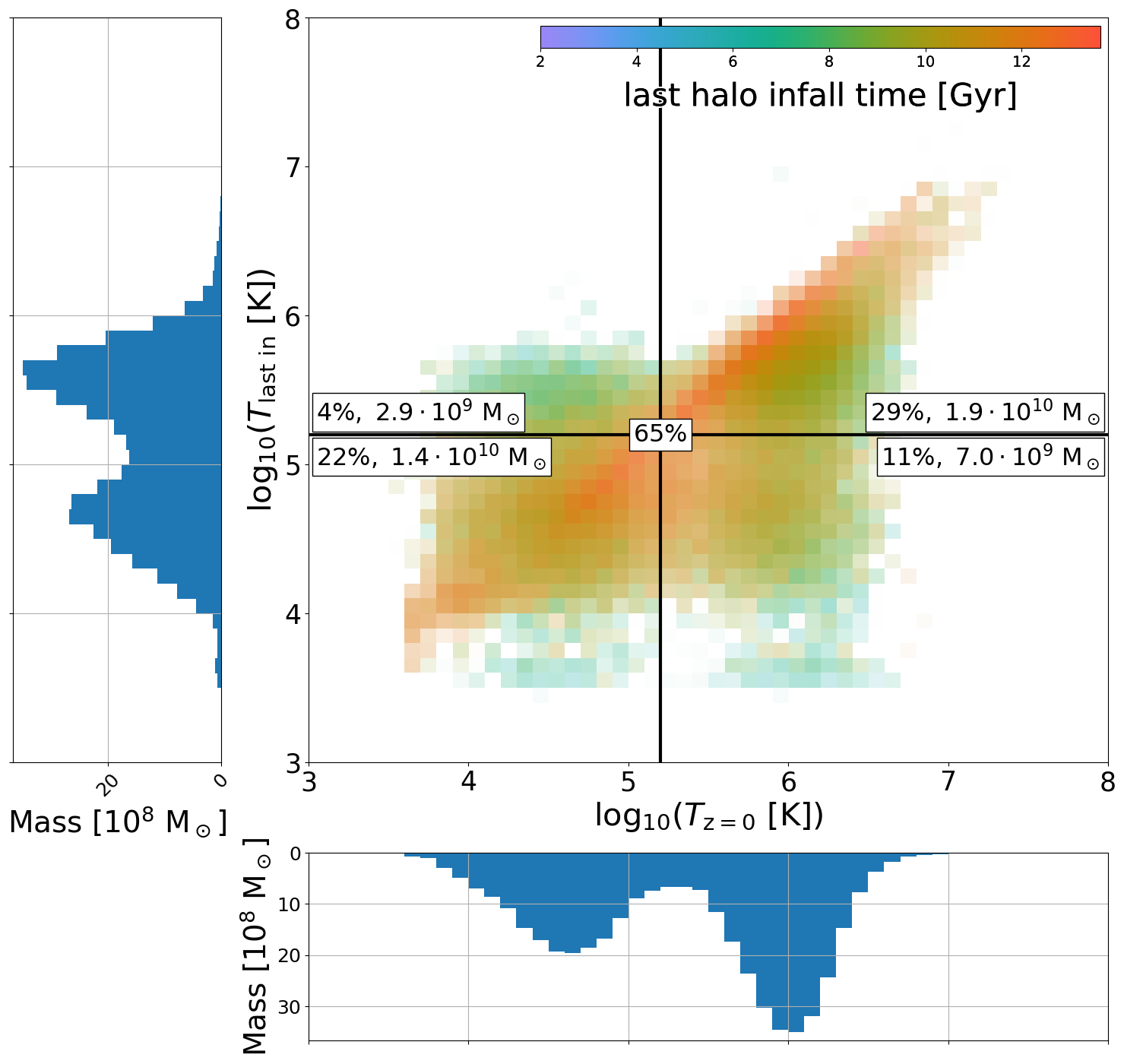}
\includegraphics[width = 0.49\textwidth]{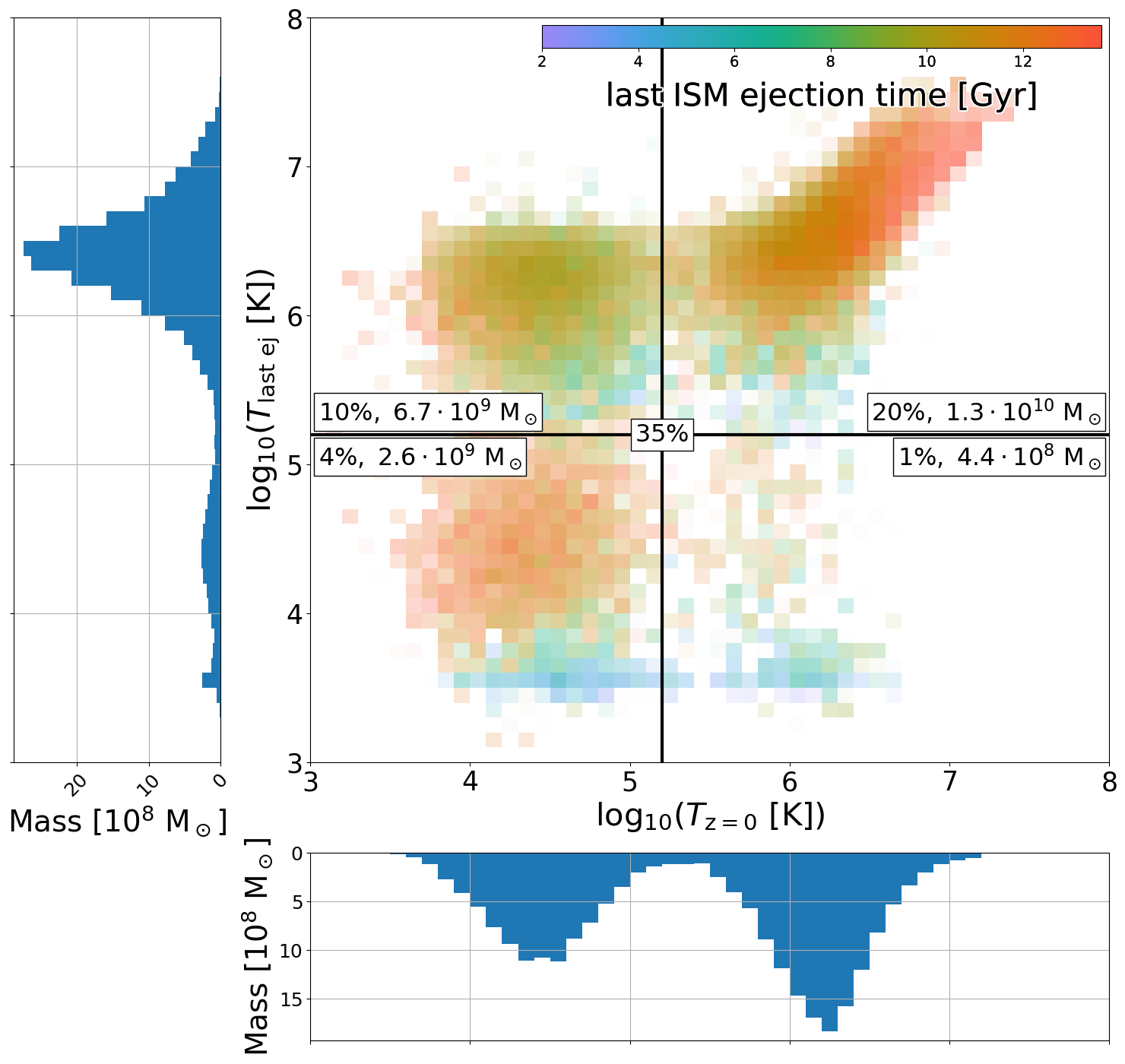}
\caption{
	The temperature distribution of the CGM gas for M08585 at $z=0$ separated into its origin. All percentages are given with respect to the total CGM mass. We have tracked the temperature of each gas phase element at infall time onto the halo ($T_{\mathrm{last\,in}}$, left) and temperature at last ejection from the galaxy into the halo ($T_{\mathrm{last\,ej}}$, right) and compare to the temperature at present $T_{\mathrm{z=0}}$. Temperature ranges are separated into cold and hot by the thermally unstable regime ($\sim 10^{5.2}\,\mathrm{K}$). Gas accreted onto the halo, which has never been part of the central galaxy (left) makes up 65\% of the CGM gas. This gas is separated into four accretion to present day transitions: hot - hot (29\%), hot - cold (4\%), cold - cold (22\%), and cold - hot (11\%). Most ($22\%+29\%$ of all CGM) of the halo accreted gas falls onto the halo within the last Gyr (left, red shaded region, see also Fig.~\ref{fig:M0858_ism_halo_ejection_hist}) and did not change its temperature. Only 4\% was accreted hot and has cooled down and 11\% was accreted cold and has been heated up. Gas ejected from the ISM of the central galaxies account for 35\% of the total CGM (right). It is separated into four last ejection ejection to present day transitions: hot - hot (20\%), hot - cold (10\%), cold - cold (5\%), and cold - hot (1\%). The ejected gas is mostly hot (30\%) and a third of it has cooled down. The ejected CGM component is also dominated by recent events (red shading, see also Fig.~\ref{fig:M0858_ism_halo_ejection_hist}). This diagram shows histories which are typical for all simulated galaxies in this paper.
} 
\label{fig:M0858_halo_infall_ism_ejection}
\end{figure*}

Figure~\ref{fig:M0858_ism_halo_ejection_hist} shows the {}{instantaneous} accretion rate into the CGM of M0858 from halo accretion ({orange}{blue}) and from material returned from the ISM ({blue}{orange}). {After the early stochastic growth phase, the h}{H}alo accretion tends to dominate the contribution to the CGM. Occasional small mergers are seen as spikes in the halo accretion rate.  {At late times, when the star formation and hence feedback strength drops, there is an increase in the contribution from winds ejected from the ISM.}{The ISM ejection rate rises with the star formation rate and both rates settle at a few solar masses per year at present} and { the halo accretion rate has dropped to a similar level. Overall halo accretion dominates the bulid-up of the CGM}. 

A more detailed view of CGM growth in M0858 - which is typical for all galaxies in our sample - is provided by examining the temperature at infall.
Figure~\ref{fig:M0858_halo_infall_ism_ejection} shows a contour plot of the temperature of the gas at the time of infall (for halo accretion) or ejection (for ISM return) on the $y$-axis, relative to the temperature in the CGM at $z=0$ along the $x$-axis.  Histograms of each are shown along the sides.  Black lines are shown at $T=10^{5.2}\,$K as a by-eye division between the cold and warm-hot phases (i.e. at the minimum in between the two histogram peaks). The percentages of total CGM gas in each quadrant are indicated as well as the respective masses.

Overall, about two-thirds (65 \%) of the gas is halo accreted, while approximately one-third (35 \%) of CGM comes from ISM ejection. The domination of halo accreted gas is typical for all our simulated systems (see $f_{\mathrm{CGM}}^{\mathrm{accreted}}$ column in Table~\ref{tab:haloprops}. This is qualitatively consistent with the very detailed tracking analysis of \citet{2019MNRAS.488.1248H} based on zoom simulations (5 zooms in our mass range) with the FIRE II feedback implementation. In general, material that entered the CGM hot remains hot and similarly for the cold component; however, about 10\% of the CGM comes from gas that was accreted cold but ended up hot, and another 10\% from mass ejected from the ISM hot but ended up cold. There is a strong tendency for hot ISM ejected gas to remain hot in the CGM; cold ISM injection tends to be subject to rapid recycling as it cannot support itself thermally against gravity.  For the halo accretion, both hot and cold accretion end up about equally contributing to the CGM, and there is only a weak trend for that gas being further heated. As a result, cold CGM gas predominantly comes from cold halo accretion, although hot ISM injection that subsequently cools provides a non-trivial contribution. These results highlight the interplay between halo heating, ISM energy injection, and subsequent CGM cooling and heating processes in setting the phase structure of the CGM.

\begin{figure}
\includegraphics[width = 0.49\textwidth]{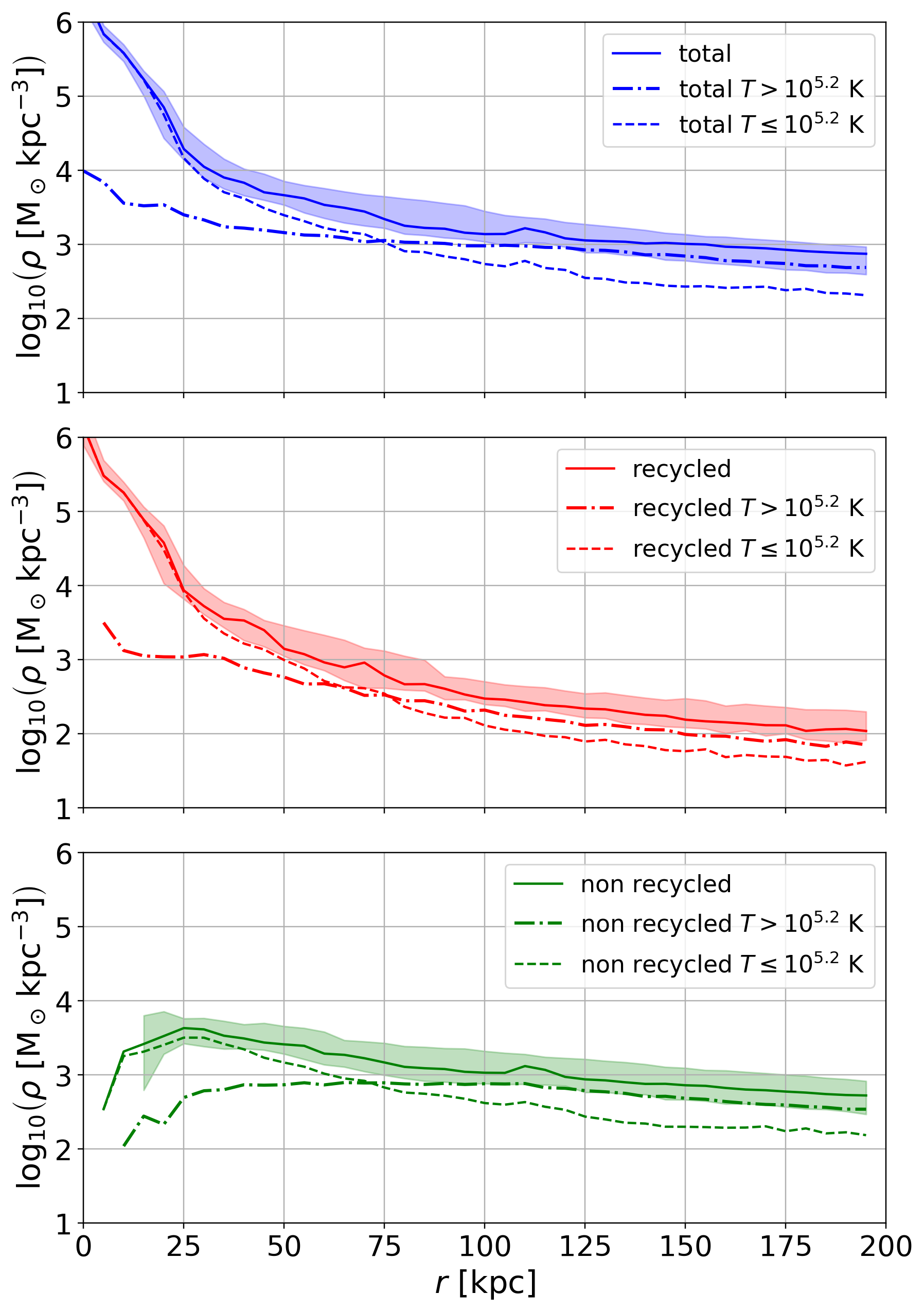}
\caption{
Average radial distribution of CGM gas for all simulated galaxies separated by temperature at $10^{5.2}\,$K at $z=0$. The top panel shows all gas, which is dominated by cooler gas at the central $75\,$kpc. In the middle and bottom panels gas is split by its origin either recycled by the central galaxy (red lines) or accreted onto the halo without ever being in the ISM throughout its past {recycling}{} (green lines). At radii larger than $50\,$kpc the CGM gas is dominated by {accreted}{non recycled} gas which might have been enriched in other galaxies before infall. 
}
\label{fig:radial_CGM}
\end{figure}

In Fig.~\ref{fig:radial_CGM} we show the average radial distribution of the CGM gas in all simulations separated in temperature below and above $10^{5.2}\,$K {}{at $z=0$}. The cooler gas (upper panel) is clearly concentrated towards the central parts ($<75\,$kpc) whereas hot gas dominates at larger radii. In the lower two panels gas is separated into recycled gas {ejected}{which in the past experienced at least one ejection}  from the central galaxies (middle panel) and non-recycled CGM gas which has only been accreted onto the halo {}{and had never been in contact with the ISM} (bottom panel). At radii larger than 50~kpc CGM {gas}{} is dominated by {}{non recycled} gas {accreted onto the halo}{}, possibly enriched and heated in infalling galaxies, but mostly unaffected by feedback events in the central galaxy. The central $50\,$kpc are dominated by recycled gas. The central dominance of recycled gas in our simulations is significantly stronger than for the simulations analysed in \citet{2019MNRAS.488.1248H}, who use a different feedback model.

\section{Mock Absorption Spectra}

We will now examine how the CGM of our simulation suite appears when observed in absorption lines. We generate mock absorption spectra. Below we describe various approaches to mock absorption spectra generation in {\sc pygad} analysis package and discuss the results. 

\subsection{Ionisation Fractions and H\textsc{i}}
\label{sec:ionisation_fracs}


To create absorption line spectra, we must calculate the ionisation fraction for the ion of interest.
Given density, temperature, and a photo-ionising background~\citep{2001cghr.confE..64H}, the ionisation fractions in the optically thin approximation can be calculated using Cloudy \citep{2013RMxAA..49..137F}. { We have not explored different estimates for the photo-ionising background in this work.}

To account for self-shielding, we follow the procedure outlined in \cite{2013MNRAS.430.2427R} to compute the effective attenuation of the ionising background in dense gas. This ionising background is then used in Cloudy to re-calculate the H\textsc{i} fraction. The fraction of molecular hydrogen (H${}_2$) is not calculated, but at the impact parameters we consider the molecular fraction is not expected to be significant.

Metal lines are also impacted by self-shielding. The same attenuation factor for the ionising background is used to recompute metal absorbers. This is not strictly accurate, as self-shielding likely changes the shape of the background, which is not accounted for.  Self-shielding tends to only affect low ions such as MgII, and even then fairly mildly.  Neutral ions like MgI would be more strongly impacted, but we do not consider such ions here since they are not typically seen in CGM absorption studies. Hence our approximation is relatively robust, and follows that used in previous such studies~\citep[e.g.][]{2016MNRAS.459.1745F}.

Table~\ref{tab:haloprops} shows the size of the H\textsc{i} disc, computed as the radius at which the face-on radial surface density profile of H\textsc{i} drops below $1\,\mathrm{M_\Sun} \mathrm{pc}^{-2}$ \citep{2014MNRAS.441.2159W}.  These values are broadly comparable to observations, but show a larger scatter than observed.

\subsection{Generating Spectra from Simulations}
\label{sec:mock_spectra}

Techniques for generating mock spectra have been developed in numerous forms~\citep[e.g.][]{2006MNRAS.373.1265O,2017ApJ...847...59H}, but none so far have combined the features of being python-based, native-SPH, and computationally efficient.  Hence we develop and implement our own spectral generation code into our publicly available {\sc pygad} package.  Along the way, we explored several variants in the way to smooth gas elements onto the line of sight (LOS), in order to better understand uncertainties associated with mock spectrum generation.

To generate a spectrum for a particular ion, we (i) compute the ionisation fraction of a given SPH particle; (ii) smooth the ion density onto the LOS via a gather approach at the LOS velocity of each pixel; (iii) convert the ion column density into an optical depth given the ion's oscillator strength.  This approach is used in e.g. the SPH-based spectra generation code \textsc{specexsnap} \citep{2008MNRAS.387..577O}.

\begin{figure*}
\centering
\includegraphics[width = 0.9\textwidth]{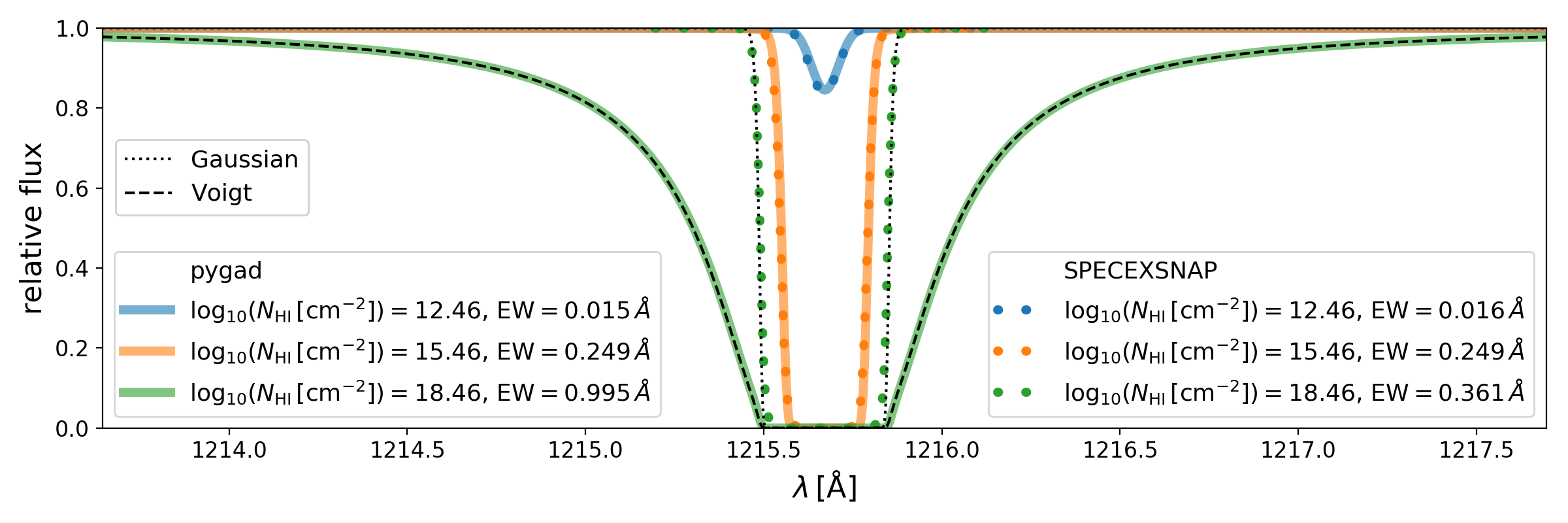}
\caption{
	Comparison of the line profile generation from \textsc{pygad} (coloured lines) and
	\textsc{specexsnap} (coloured dots) from an idealised setup containing a cold blob of homogeneous gas with varying column densities (different colours) in a hot environment. The difference between a Gaussian profile (\textsc{specexsnap}, black dotted) and a Voigt profile (\textsc{pygad}, dashed lines) becomes clearly visible at the highest column density (green).
}
\label{fig:specexsnap}
\end{figure*}

To test this implementation and compare to \textsc{specexsnap}, we set up homogeneous spheres of cool dense \mbox{($T = 10^4\,\mathrm{K},\ n_H = 0.1\,\mathrm{cm}^{-3}$)} gas at fixed density and temperature embedded in hot low density \mbox{($T = 10^6\,\mathrm{K},\ n_H = 0.001\,\mathrm{cm}^{-3}$)} ambient medium. The size of the sphere is varied to result in hydrogen column densities of $10^{13}\,\mathrm{cm}^{-2}$, $10^{16}\,\mathrm{cm}^{-2}$, and $10^{19}\,\mathrm{cm}^{-2}$, respectively.
Due to partially ionised hydrogen, the H\textsc{i} column densities are about $0.5$~dex lower. These cases cover the linear, {exponential}{logarithmic}, and square-root regimes of the curve of growth for H\textsc{i} (see section \ref{sec:ewcd}). In Fig. \ref{fig:specexsnap} we compare the two codes. For the two low column density cases (blue and orange) the results are identical. At the highest column density the natural line width dominates the profile which is perfectly captured by \textsc{pygad} when compared to the theoretical Voigt profile (black dashed line). \textsc{specexsnap} is limited to Gaussian line generation (black dotted line). 

Another approach is to smooth the physical quantities onto the LOS first, and then compute the ionisation fraction from those smoothed physical quantities (e.g. binning data on a grid).  In effect this switches steps (i) and (ii) in the above procedure.  This is generally not preferred because particles with significantly different physical conditions (e.g. temperatures) can legitimately be regarded as representing multi-phase gas, in particular in the simulation approach followed here.  Averaging the physical properties removes this multi-phase nature, and creates ionisation fractions that reflect an average of multiple phases. To compare these approaches, we have implemented both methods into {\sc pygad}.

\begin{figure*}
\centering
\includegraphics[width = 0.9\textwidth]{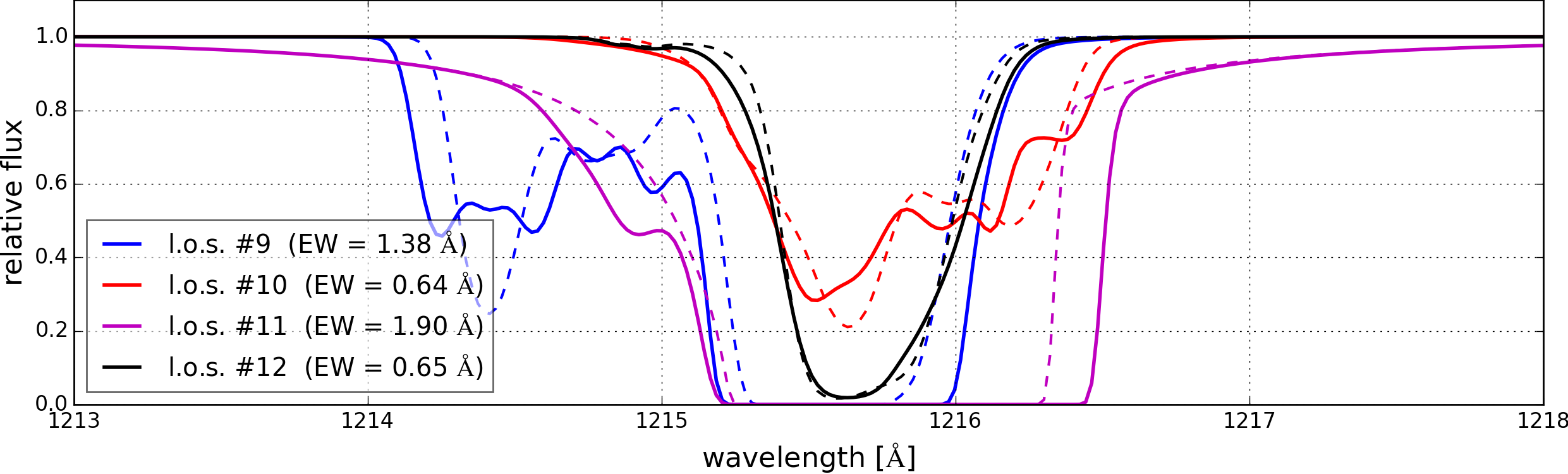}
\caption{
    Mock Lyman-$\alpha$ absorption spectra of M0858 for lines \#9 through \#12,
    cf.\ Fig.~\ref{fig:M0858_H1215_maps_zoom}.
    Solid lines are generated by our fiducial method and the dashed lines
    by the smoothing first approach (cf.\ section~\ref{sec:mock_spectra}).
    Typically, the fiducial method
    results in boarder lines in the saturated regions and more substructure in the
    line wings.
    The EWs (\#9 through \#12) are $1.38$~\AA\ ($1.14$~\AA), $0.64$~\AA\
    ($0.62$~\AA), $1.47$~\AA\ ($1.14$~\AA), and $0.65$~\AA\ ($0.61$~\AA) for the
    fiducial (smoothing first) method.
    For line~\#11 we can see damping wings.
}
\label{fig:M0858_Lya_line_profiles_special}
\end{figure*}

\begin{figure*}
\centering
    \includegraphics[width = 0.78\textwidth]{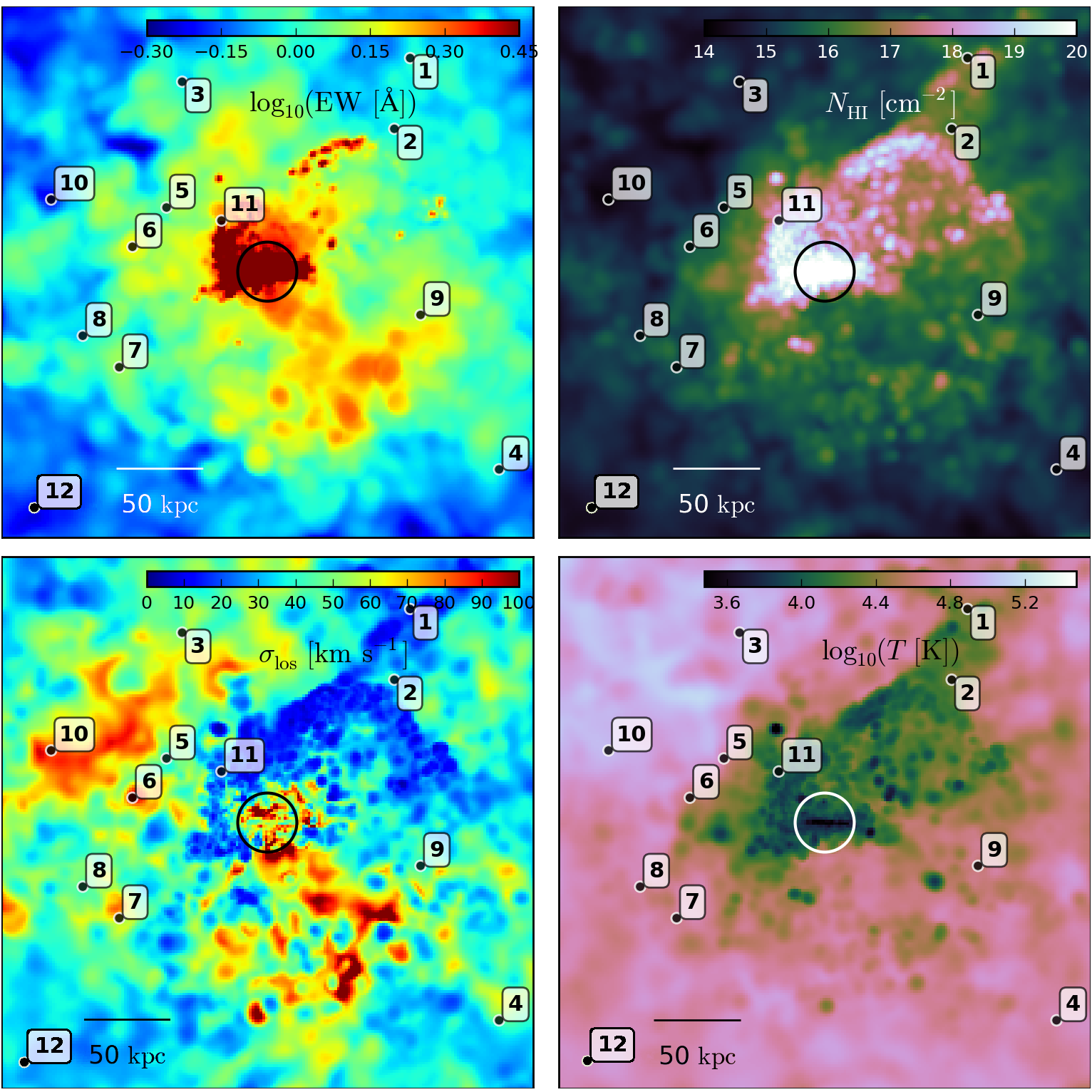}
\caption{
	Projection of M0858 in a region $306\,\mathrm{kpc}$ on a 
	side. We show the Lyman-$\alpha$ EWs (upper left), column densities of H\textsc{i} (upper right), , the 
	H\textsc{i} gas velocity dispersion (bottom left), and the mean H\textsc{i} mass weighted gas 
	temperature (bottom right).
	EWs in the densest regions ($\gtrsim10^{15}\,\mathrm{cm}^{-2}$, logarithmic regime) do not change 
	significantly with column density, but mostly with LOS velocity structure. The sample spectra of Fig.~\ref{fig:M0858_Lya_line_profiles_special}, 
	\ref{fig:M0858_Lya_line_profiles_EW1}, and \ref{fig:M0858_Lya_line_profiles_N155} are indicated with numbers next to them.}	
\label{fig:M0858_H1215_maps_zoom}
\end{figure*}

Figure~\ref{fig:M0858_Lya_line_profiles_special} shows example absorption Lyman-$\alpha$ spectra of halo M0858 for four selected lines-of-sight with both methods (for the location of each LOS see Fig.~\ref{fig:M0858_H1215_maps_zoom}).
The solid spectra are generated with our fiducial method, while
the dashed spectra are generated by first smoothing the physical quantities.
The general trend is that  the fiducial method shows larger
equivalent widths, often by 20\% or more, possibly due to a better representation of the broadening by LOS velocity.
These spectra were chosen to emphasise the difference between the methods, especially lines \#9 and \#10.
More typical spectra are shown in Figures ~\ref{fig:M0858_Lya_line_profiles_EW1} and
\ref{fig:M0858_Lya_line_profiles_N155}.
Line \#11, is a common example, where an additional absorption feature
occurs next to the main feature and causes it to become significantly wider.  This shows that, in some cases, the choice of exactly how to generate mock spectra can result in mild but nonzero differences in the resulting column densities.

\section{Halo Absorption Properties}

\begin{table}
\centering
\begin{tabular}{l|cccS}
    number & $\log_{10}(N\,[\mathrm{cm}^{-2}])$ & $EW$ [\AA ] & $\sigma\,[\mathrm{km/s}]$ & $\rho\,[\mathrm{kpc}]$ \\
    \hline
    1  & 16.46 & 0.982 & 19.5 & 147.3 \\
    2  & 16.35 & 0.968 & 33.5 & 110.0 \\
    3  & 14.72 & 0.903 & 51.3 & 120.1 \\
    4  & 14.85 & 0.948 & 44.8 & 175.5 \\
    5  & 15.55 & 1.136 & 41.7 &  68.1 \\
    6  & 15.64 & 1.535 & 92.9 &  79.0 \\
    7  & 15.55 & 1.278 & 63.9 & 101.3 \\
    8  & 15.34 & 0.889 & 44.6 & 112.8 \\
    9  & 15.96 & 1.379 & 38.7 &  92.0 \\
    10 & 14.22 & 0.636 & 81.4 & 130.3 \\
    11 & 18.66 & 1.897 & 22.7 &  40.3 \\
    12 & 14.47 & 0.651 & 30.3 & 185.0
\end{tabular}
\caption{
	The properties of the chosen lines:
	the column density $N$, the equivalent width $EW$, velocity dispersion $\sigma$, and impact parameter $\rho$.
	The spectra are shown in Fig.~\ref{fig:M0858_Lya_line_profiles_special},
	\ref{fig:M0858_Lya_line_profiles_EW1}, and 
	\ref{fig:M0858_Lya_line_profiles_N155} and the actual positions are indicated
	in Fig.~\ref{fig:M0858_H1215_maps_zoom}.}
\label{tab:lineprops}
\end{table}

We now examine the absorption properties of CGM gas in our zoom suite.  While we will compute statistics from our full sample, Halo M0858 will continue to serve as an example throughout the work, since it is fairly typical and contains a variety of interesting features.
\begin{figure*}
\centering
\includegraphics[width = 0.9\textwidth]{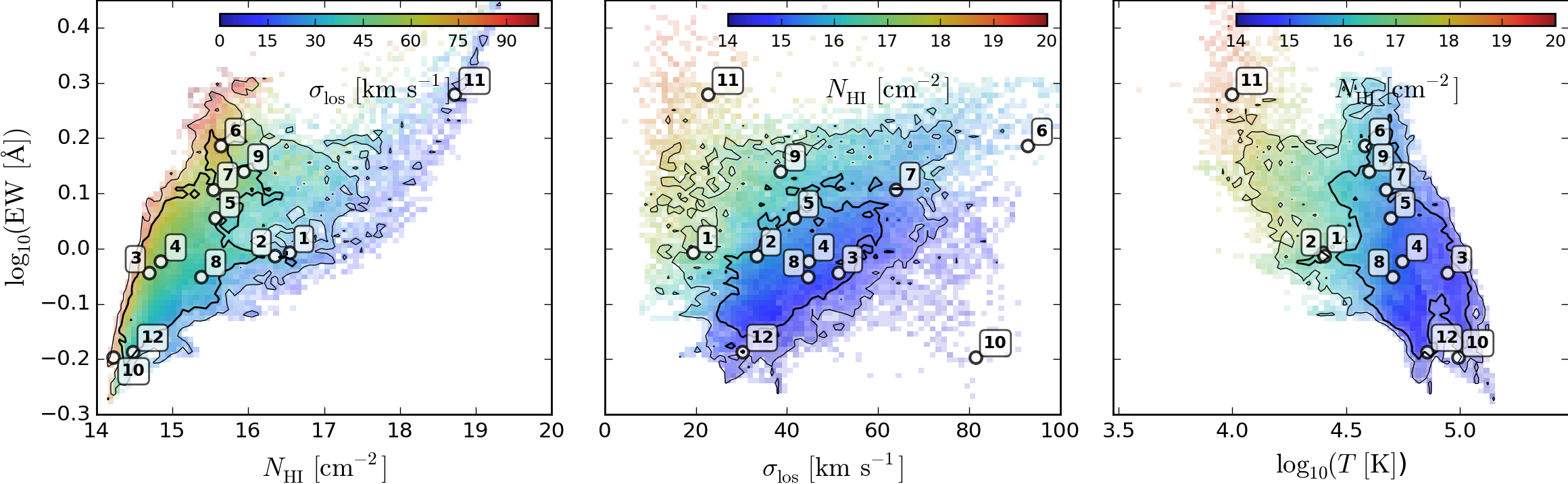}
\caption{Correlations of EW with column density, line of sight dispersion, and temperature for the maps in  Fig.~\ref{fig:M0858_H1215_maps_zoom} Special lines-of-sight are numbered and discussed in the text.}
\label{fig:M0858_H1215_corr_zoom}
\end{figure*}
\subsection{CGM Absorption Structure}

\begin{figure*}
\centering
\includegraphics[width = 0.9\textwidth]{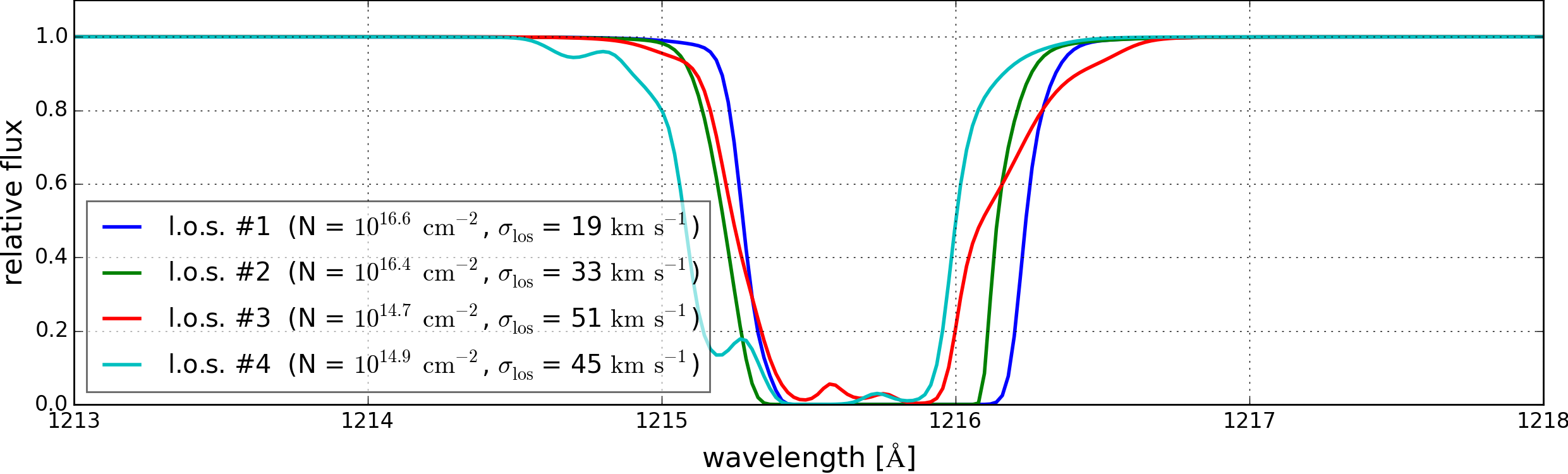}
\caption{
    Lyman-$\alpha$ absorption spectra for lines \#1 through \#4 as marked in
    Fig.~\ref{fig:M0858_H1215_maps_zoom} (cf.\ also
    Fig.~\ref{fig:M0858_H1215_corr_zoom}).
    The lines have similar EWs $\sim 0.95$~\AA, but different LOS
    velocity dispersions.
    In general the broadening due to line of sight dispersion/velocity effects 
    cannot be identified as the lines are saturated. For a few cases, like line~\#4, weak separate lines can appear (at
    $\sim\!1215.1$~\AA).
}
\label{fig:M0858_Lya_line_profiles_EW1}
\end{figure*}
\begin{figure*}
\centering
\includegraphics[width = 0.9\textwidth]{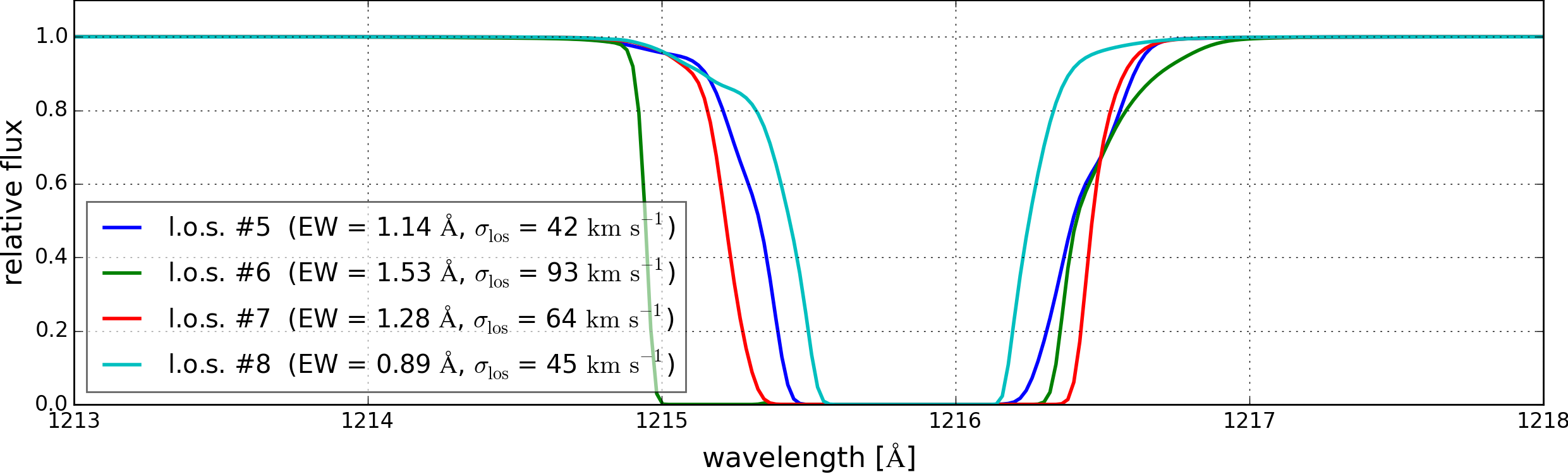}
\caption{
    Lyman-$\alpha$ absorption spectra for lines \#5 through \#8 as marked in
    Fig.~\ref{fig:M0858_H1215_maps_zoom} (cf.\ also
    Fig.~\ref{fig:M0858_H1215_corr_zoom}).
    All lines have similar column densities of about
    $10^{15.6}\,\mathrm{cm}^{-2}$, but varying EWs due to their
    different LOS velocity structure.
}
\label{fig:M0858_Lya_line_profiles_N155}
\end{figure*}

For halo M0858 we create a regular grid of $200\times200$ equally spaced sight
lines along the $x$-axis. For each sight line we generate a mock Lyman-$\alpha$
spectrum with our fiducial method and calculate the corresponding equivalent width.
Figure~\ref{fig:M0858_H1215_maps_zoom} shows the 306~kpc per side map of EW (top left), H\textsc{i} column
densities (top right), the H\textsc{i} LOS velocity dispersion (bottom left), and the
mean H\textsc{i}-mass weighted temperatures (bottom right).

On large scales, the structure of the EW map is well correlated with the column densities. There is a dense central region hosting the main galaxy, an accretion flow from the upper right that is evident as a low-temperature, low-dispersion extension, and an outflow above and below the (horizontal) disk as particularly evident from the larger velocity dispersion.


Looking more closely, discrepancies between the equivalent widths and the column densities emerge, particularly in the regions of the inflow and outflow.  While the densest knots are apparent in both high column and EW, absorption at $N_{HI}\sim\!10^{18}\,\mathrm{cm}^{-2}$ particularly along the accreting filament are not obvious as high-EW structures.  Conversely, the outflow to the lower right shows higher EWs, with no corresponding features in the column density.  This indicates local dynamical effects can complicate the association between column density and EW.

To explore this more quantitatively, we show in Figure~\ref{fig:M0858_H1215_corr_zoom} scatter plots of EW vs. $N_{HI}$ (left panel), the mass weighted line-of-sight velocity dispersion $\sigma_{los}$ (middle), and the temperature $T$ (right), for all the lines of sight.  In the left panel the points are colour-coded by the $\sigma_{los}$, while in the other two panels they are colour coded by $N_{HI}$.

While there is a general trend of higher EW corresponding to higher $N_{HI}$, there is a substantial scatter in EW at a given column density.  Moreover, this scatter correlates well with $\sigma_{los}$. This is not surprising, since saturated lines can be broadened by both enhanced columns and by enhanced velocity broadening.  However, it is notable that at the highest columns, the trend is fairly tight with EW, as the dispersion tends to be quite low along these LOS.

The low dispersion at high columns is also evident in the middle panel, where the strongest lines appear at the lowest $\sigma_{los}\approx 10-20$~km/s.  For the weaker lines there is no strong correlation between absorption strength and LOS dispersion, though there is a weak trend that the highest dispersions create somewhat larger EWs, as might be expected from line broadening.

The strong lines with low $\sigma_{los}$ also correspond to gas with the lowest temperatures, typically $T\approx 10^4$K. Since this corresponds to a Gaussian velocity dispersion of around 9~km/s, it explains why no $\sigma_{los}$ values are seen below this. Moving to the weaker absorbers, the temperatures tend to be higher. { In section \ref{sec:ewcd} we discuss in detail how the CGM motion along the line-of-sight affects the equivalent width, in addition to the thermal line broadening.}

For further analysis we marked lines with similar EW (\mbox{$\sim\!0.95$~\AA}) and column densities varying by about two orders of magnitude as \#1 through \#4  (see Figures ~\ref{fig:M0858_H1215_maps_zoom} and \ref{fig:M0858_H1215_corr_zoom}). The line profiles are shown in  Fig.~\ref{fig:M0858_Lya_line_profiles_EW1}. Naively, higher column densities would result in higher EWs. 
However, this trend is compensated by lower temperatures and lower LOS velocity dispersion (see middle and right panels of Fig.~\ref{fig:M0858_H1215_maps_zoom}). EW increase due to higher column density is compensated by a lower $b$-parameter and a lower LOS dispersion {}{resulting in lines \#1 - \#4 lining up almost horizontally in the left panel of Fig.~ \ref{fig:M0858_H1215_corr_zoom}}.

Lines-of-sight with very similar column densities of about $\sim10^{15.6}\,\mathrm{cm}^{-2}$ but strongly varying EWs are highlighted in \#5 through \#8 (Fig.~\ref{fig:M0858_Lya_line_profiles_N155}).
Line \#8 has the lowest EW of $0.90$~\AA\ with a velocity dispersion of
\mbox{$\sim\!45\,\mathrm{km}\,\mathrm{s}^{-1}$}, whereas line \#6 has an
equivalent width of $1.53$~\AA\ and a $\sigma_\mathrm{los}\sim\!100\,\mathrm{km}\,\mathrm{s}^{-1}$.
This is an example where higher velocity dispersion generates larger EW at similar column density{}{, with lines \#5 - \#8 lining up vertically in the left panel of Fig.~ \ref{fig:M0858_H1215_corr_zoom}}.

\subsection{Individual Line Profiles}

The gas velocity structure along each LOS can play an important role for
the determination of EWs at the disc-halo interface, where the Lyman-$\alpha$ column densities are in the logarithmic regime of the curve of growth.
In our simulations this regime stretches out to a few H\textsc{i} disc sizes, almost reaching the $R_{200}$ of our galaxies.  Here we examine some individual LOS to see how they reflect the underlying CGM physical conditions.

The line spectra of the already discussed sight lines are plotted in
Fig.~\ref{fig:M0858_Lya_line_profiles_EW1} (\#1 - \#4),
Fig.~\ref{fig:M0858_Lya_line_profiles_N155} (\#5 - \#8), and
Fig.~\ref{fig:M0858_Lya_line_profiles_special} (\#9 - \#12).
The lines have impact parameters between $2\,R_\mathrm{HI}$ and
$11\,R_\mathrm{HI}$ at H\textsc{i} column densities of $10^{14}\,\mathrm{cm}^{-2}$
to $10^{17}\,\mathrm{cm}^{-2}$, and one at $10^{19}\,\mathrm{cm}^{-2}$.
Most lines are saturated at EWs higher than $0.8$~\AA,
the only exceptions being lines~\#10 and \#12 with EWs of $\sim\!0.6$~\AA.
Due to the saturation, the Lyman-$\alpha$ absorption lines do not show clear internal structure. 
This makes it impossible to distinguish between line broadening by column density, temperature $b$-parameter (line~\#3) or line-of-sight velocity structure (line~\#2).

\begin{figure}
\centering
\includegraphics[width = 0.48\textwidth]{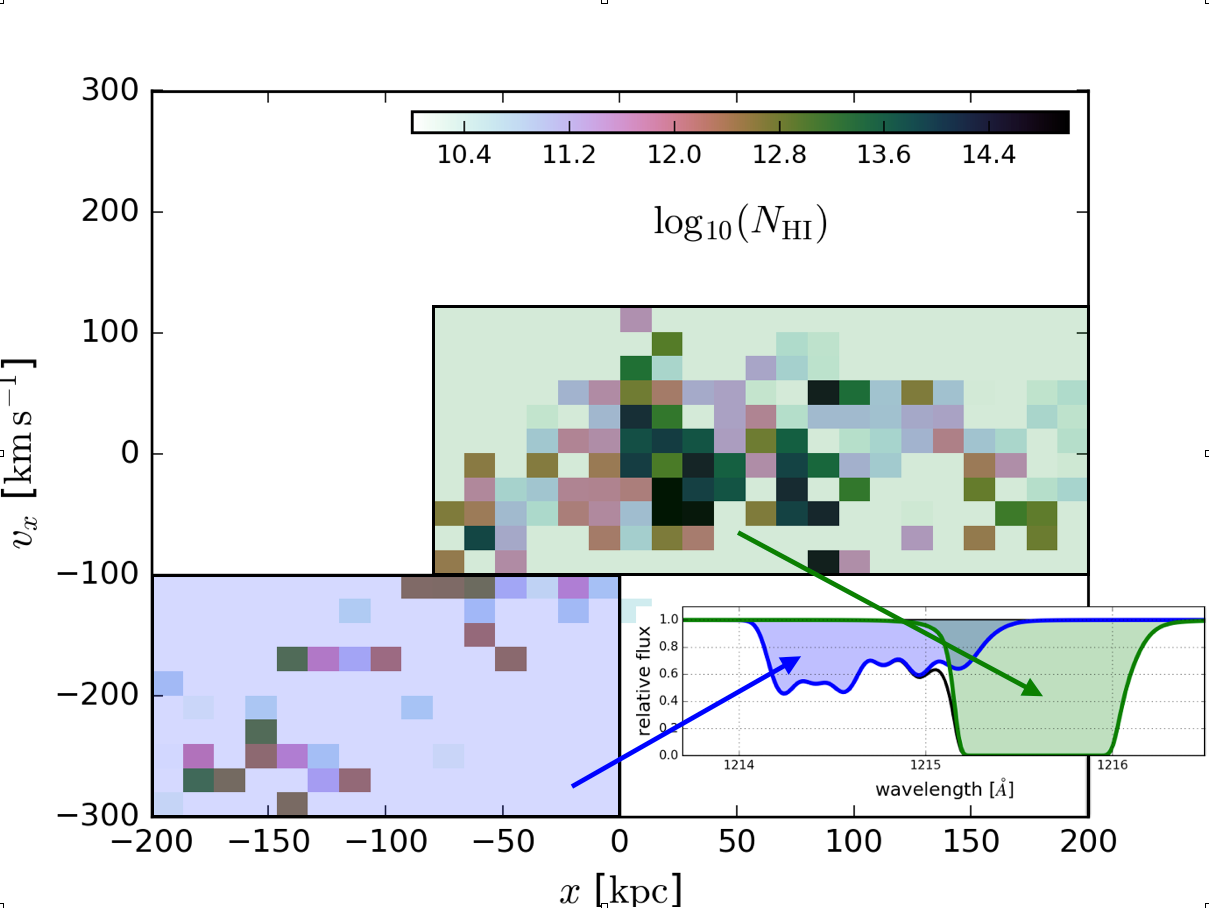}
\caption{
	The column density distribution along the line of sight number~9 (cf.\ Fig.~\ref{fig:M0858_Lya_line_profiles_special}, \ref{fig:M0858_H1215_maps_zoom}, and \ref{fig:M0858_H1215_corr_zoom}) and as a function of its velocity $v_x$ and position $x$ along the line of sight.
	In the 2D-histogram the amount of column density in each pixel is colour coded.
	In the inlayed spectrum of the line (cf.\ Fig.~\ref{fig:M0858_Lya_line_profiles_special}) we separated the the H\textsc{i} mass by its position and velocity as indicated by the coloured regions in the phase space and created spectra of just that mass.
	The mass extending towards negative positions and velocities (light blue square) created the left wing (also light blue) in the spectrum, whereas the rest creates the main line.} 
\label{fig:vlos_vs_xlos_l9}  
\end{figure}

Only in a few cases (lines~\#9 and \#10) do we see more complex line shapes, often generated by several resolved lines. { A solid physical interpretation of these structures from simulations in general is difficult, both because the interpretation can change with the method to generate the lines. Here we highlight one line which is not saturated and allows for a detailed physical interpretation of the complex line shape. While this cannot be generalised it nevertheless indicates how simulations can support the analysis of complex line shapes.} 
In Fig.~\ref{fig:vlos_vs_xlos_l9} we show the distribution of the H\textsc{i} column density in LOS velocity along the sight line~\#9 (x-axis), as an example. 
It covers $400\,\mathrm{kpc}$ and about $400\,\mathrm{km}\,\mathrm{s}^{-1}$ in velocity space.
Alongside the total spectrum we show spectra of gas in the marked (blue and green) regions artificially separated in velocity space. 
Most of the mass is at line-of-sight positions between $x=0\,\mathrm{kpc}$ and $x=100\,\mathrm{kpc}$ with absolute velocities of less than $100\,\mathrm{km}\,\mathrm{s}^{-1}$.
This corresponds to a total shift of about $0.4$~\AA, a bit less than its EW.
Towards negative positions we find gas with line-of-sight velocities down to $\sim\!300\,\mathrm{km}\,\mathrm{s}^{-1}$, explaining the $\sim\!1$~\AA\ wide wing in the spectrum towards shorter wavelengths. 

\subsection{Radial Equivalent Width Profiles}

\begin{figure}
\centering
\includegraphics[width=\columnwidth]{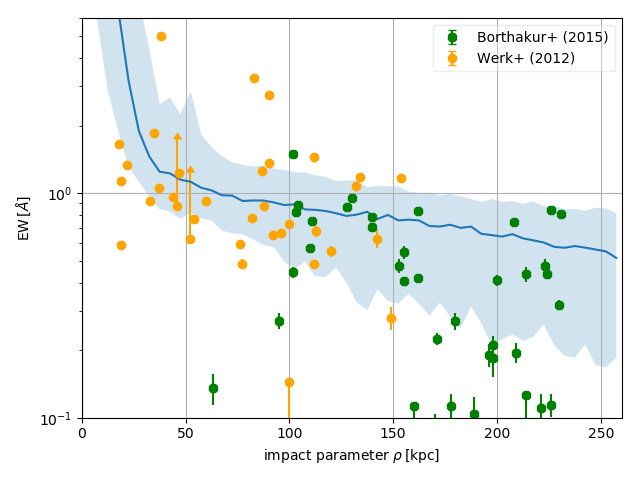}
\caption{
    The equivalent widths of the Lyman-$\alpha$ lines for all halos in edge-on projection as a function of their impact radius $\rho$.
	The green points are observational from \protect\cite{2015ApJ...813...46B} and the orange from \protect\cite{2013ApJ...777...59T}.
	} 
\label{fig:EW_profile_all}
\end{figure}

By combining single LOS observations around a suite of galaxies, it is possible to characterise the typical absorption profile in the CGM.  As an example, the COS-GASS survey~\citep{2015ApJ...813...46B} selected galaxies having H\textsc{i} data, and found a clear correlation of more gas-rich galaxies (as seen in 21$\, $cm emission) having higher H\textsc{i} EW absorption.  Here we compare our suite of simulations to the H\textsc{i} absorption as a function of impact parameter from these data.

Figure~\ref{fig:EW_profile_all} shows the average radial H\textsc{i} EW profiles of all our seven galaxies without a (major) merger (see also Table~\ref{tab:haloprops}) out to several H\textsc{i} disc sizes. {The M0858 halo is plotted to larger radii.}{} Overplotted are observations from the GASS, GBT, CHA, and ALFALFA surveys as taken from \cite{2015ApJ...813...46B}.

The simulated radial profiles follow the observed trends reasonably well, both in amplitude and trend. This is in agreement with results from \citet{2017MNRAS.464.2796G} who compare H\textsc{i} surface density profiles from the NIHAO zoom simulations to H\textsc{i} column densities inferred from simulations. There is a lack of low-EW absorption seen in our simulations, indicating somewhat more absorption than typically observed.  A partial explanation might be found in survey selection, as the observations have a median stellar mass of $10^{10.4}\,\mathrm{M_\Sun}$, H\textsc{i} mass of $10^{9.4}\,\mathrm{M_\Sun}$, and disc size $R_\mathrm{HI}$ of $16.2\,\mathrm{kpc}$, whereas our sample chosen solely by mass has a median stellar mass of $10^{10.8}\,\mathrm{M_\Sun}$, a H\textsc{i} mass of $10^{10.0}\,\mathrm{M_\Sun}$ and a disc size $R_\mathrm{HI}$ of $25\,\mathrm{kpc}$.
Hence our simulated galaxies are somewhat larger, potentially giving rise to higher EW.  Nonetheless, the general predicted trend of a broad range of CGM absorption strengths as a function of impact parameter is broadly consistent with observations.

\subsection{EW vs. Column Density Profiles}\label{sec:ewcd}

\begin{figure}
\centering
\includegraphics[width=\columnwidth]{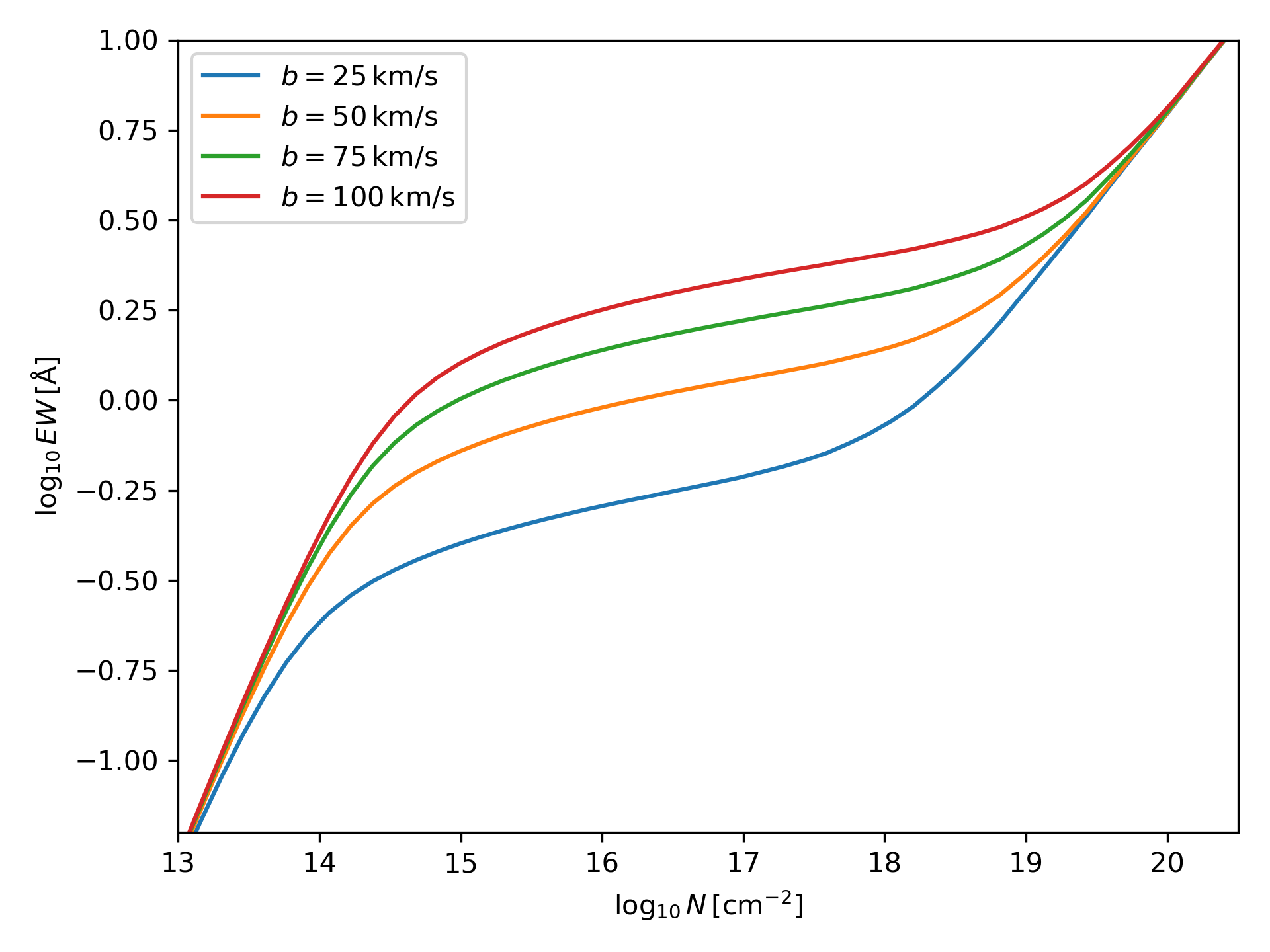}
\caption{Four example curves of growth (COG) for Lyman-$\alpha$ with $b$-parameters of {20}{25}, 50, 75, and 100 km/s.
}
\label{fig:multiple_b}
\end{figure}

In Figure~\ref{fig:multiple_b} we show four examples for curves of growth (COG) relating column density to EW for Lyman-$\alpha$ with $b$-parameters of {20}{25}, 50, 75, and $ 100\,\mathrm{km}\,\mathrm{s}^{-1}$. In the linear regime at low column densities ($\log N_{HI} \lesssim 14$) the line is optically thin. At higher column densities the line becomes saturated (linear or logarithmic regime), and at higher column densities it is dominated by the outer Lorentzian wings of the natural line broadening (square-root regime at $\log N_{HI} \gtrsim 19$). Most CGM Lyman-$\alpha$ absorbers are in the logarithmic regime of the curve of growth where changes in column density do not affect EWs strongly. As the thermal $b$-parameter for $T = 10^4\,\mathrm{K}$ gas is only $ \sim 13\,\mathrm{km}\,\mathrm{s}^{-1}$ we expect the broadening to be dominated by CGM kinematics. By comparing the EW and column density radial profiles from the simulations we can derive an effective $b$-parameter which might be useful to convert observed Lyman-$\alpha$ EWs to column densities.

\begin{figure*}
\centering
\includegraphics[width=0.49\textwidth]{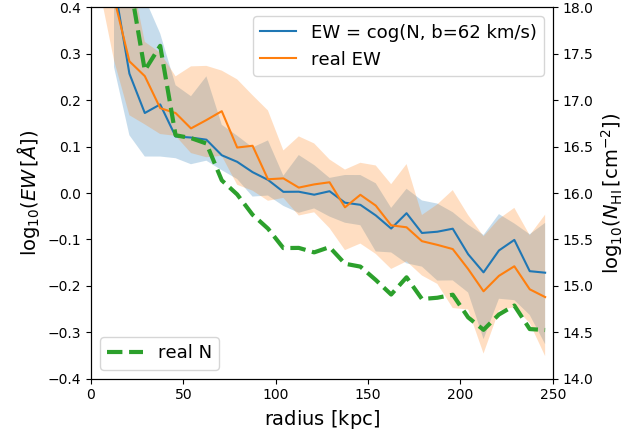}
\includegraphics[width=0.49\textwidth]{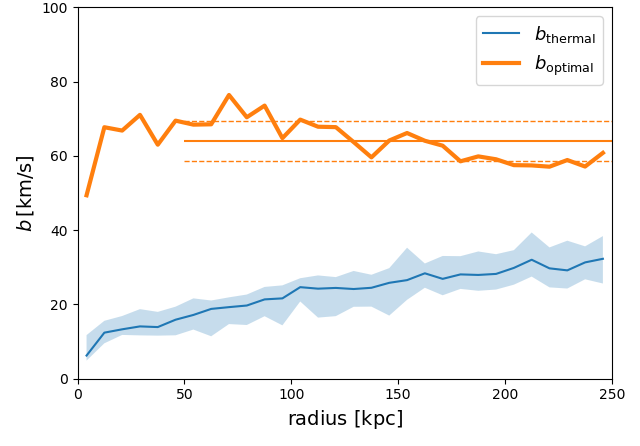}
\caption{
	Radial profiles of EW and column density for M0858 in edge-on projection on the left with arbitrary matching of the scales.
	EWs as obtained by the mock line spectra (orange line) with the shaded region indicating the 10th and 90th percentile.
	The simulated column densities (green line), omitting the scatter for better visibility (note: full transition into logarithmic region is at $\sim\!10^{15}\,\mathrm{cm}^{-2}$).
	EW calculated from the LOS column densities and assuming a COG with $b = 62\,\mathrm{km/s}$ (blue line) with the shaded region indicating the 10th and 90th percentile.
	\newline
	The right panel shows the thermal $b$-parameters as a function of radius (blue line) and the `optimal' $b$-parameters (orange line) for M0858.
	This `optimal' $b$-parameter is the value for which the rms of difference between the (logarithmic) EWs obtained from a COG and the actual EWs is the smallest.
	The solid horizontal line indicates the mean over this radial range and the dashed lines the rms.
}
\label{fig:M0858_EW_N_profile_optimal_b}
\end{figure*}

\begin{figure}
\centering
\includegraphics[width=0.5\textwidth]{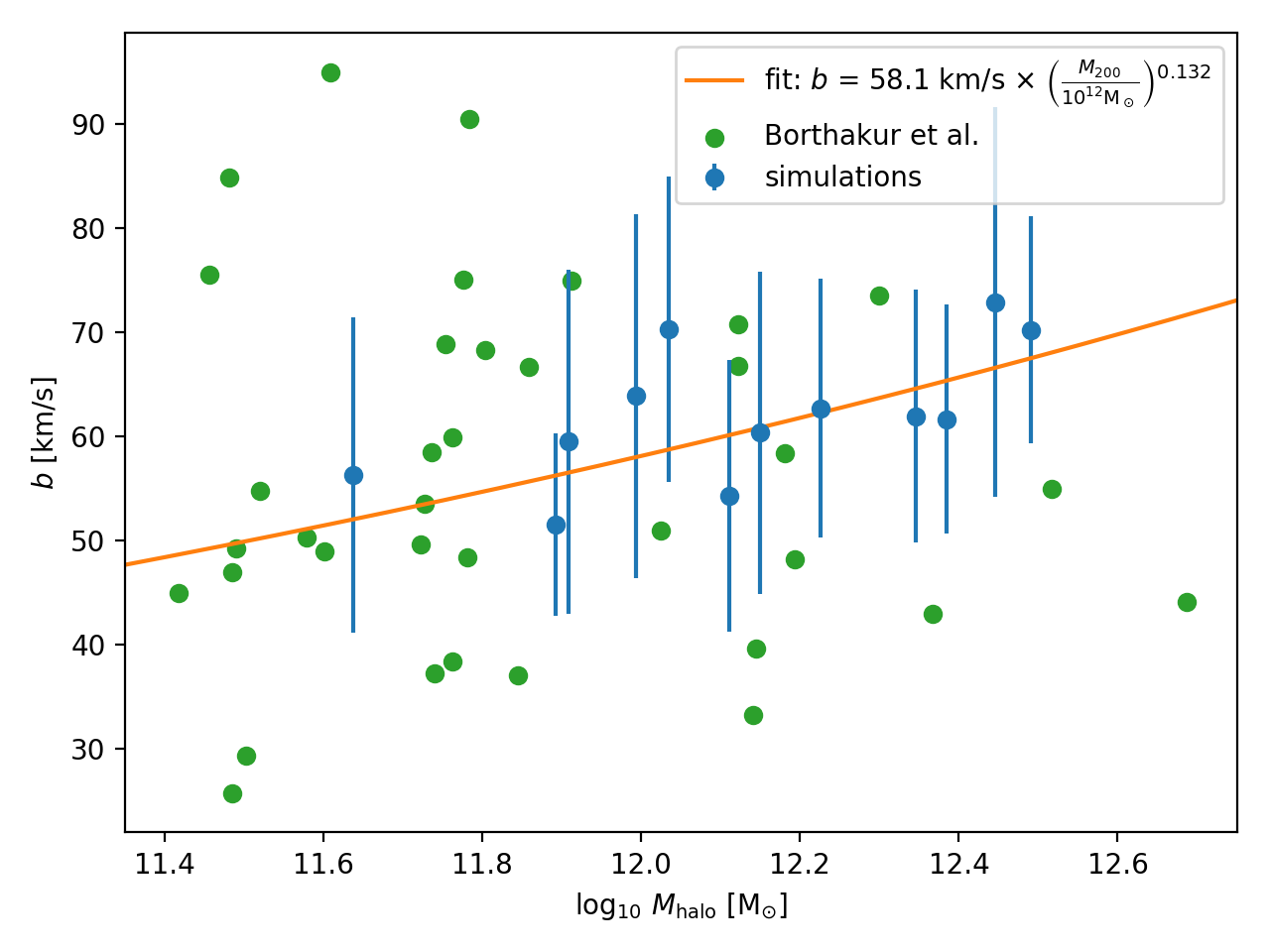}
\caption{
	The `optimal' $b$-parameters (cf.\ Fig.~\protect\ref{fig:M0858_EW_N_profile_optimal_b} and text) over the radial range $50\,\mathrm{kpc}$ to $250\,\mathrm{kpc}$ in edge-on projections as a function of halo mass for all simulated galaxies.
	The error bars show the standard deviations of the optimal $b$-parameters.
	The data was fitted with a an exponential function: \mbox{$b = b_0 \left[ M_{200} / (10^{12}\,\mathrm{M_\Sun}) \right]^x$} with a best fit of \mbox{$b_0 = 58.1\,\mathrm{km/s}$} and $x=0.132$.
	For M0858 ($M_\text{halo} = 10^{12.23}\,\mathrm{M_\Sun}$) this results in $b = 62\,\mathrm{km/s}$ as used in Fig.~\protect\ref{fig:M0858_EW_N_profile_optimal_b}. Data taken from \citet{2016ApJ...833..259B}.
}
\label{fig:optimal_b}
\end{figure}

In the left panel of Fig.~\ref{fig:M0858_EW_N_profile_optimal_b} we show the simulated column density distribution (green dashed line) and the EW distribution derived with \textsc{pygad} {}{by including all absorption along the line of sight} (orange line). A remarkably good match (blue line) to the this EW distribution can be obtained assuming a single COG with an effective $b$-parameter of $b=62$~km/s when converting column densities to EWs. In the right panel of Figure~\ref{fig:M0858_EW_N_profile_optimal_b} we show that such an optimal $b$-parameter is almost independent of radius (orange line). Here the $b$-parameter has been determined at each radial distance from finding the optimal COG (see Figure~\ref{fig:multiple_b}) resulting in the smallest rms difference between the COG inferred EW and the EW from the \textsc{pygad} analysis. The optimal $b$-parameter is clearly higher than the Doppler parameter for simple thermal broadening of hydrogen $b = \sqrt{k_B T/m_{\mathrm{ion}}}$ (blue line in Figure\ref{fig:multiple_b}).
This indicates that velocity broadening is dominating the EWs of these absorbers.

We conduct an analogous analysis for all other halos. They consistently show an optimal $b$-parameter that has very little variation with radius. However, the values of the $b$-parameters change with halo mass and for lower mass halos the contribution from the thermal broadening is more significant.

Figure~\ref{fig:optimal_b} shows the mean optimal $b$-parameter with standard deviation in the radial range from $50\,\mathrm{kpc}$ to $250\,\mathrm{kpc}$ for all halos as a function of their mass.
We see a clear trend of the optimal $b$-parameter increasing with halo mass. The correlation is worse with stellar mass, gas mass, or baryonic mass.
The exponential fit to the data is:
\begin{align}
b_\text{optimal} = \left( \frac{M_{200}}{10^{12}\,M_\Sun} \right)^{0.132} \,  58.1\,\mathrm{km/s}.
\label{eq:optimal_b}
\end{align}
We note that low-redshift observational CGM studies have typically assumed $b$ parameters in the range of 25--30~km/s when converting from EW to column density~\citep[e.g.][]{2014ApJ...792....8W}.  This is generally well below the values suggested by our simulations. Hence the observationally inferred column densities may be significantly overestimated, which could also affect the inferred halo baryon budget. We would like to point out that observations indicating higher $b$ parameters \citep[see][Fig. 6]{2016ApJ...833..259B} are consistent with our findings and a trend with halo mass is not seen.

\begin{figure*}
\centering
\includegraphics[height = 0.25\textwidth]{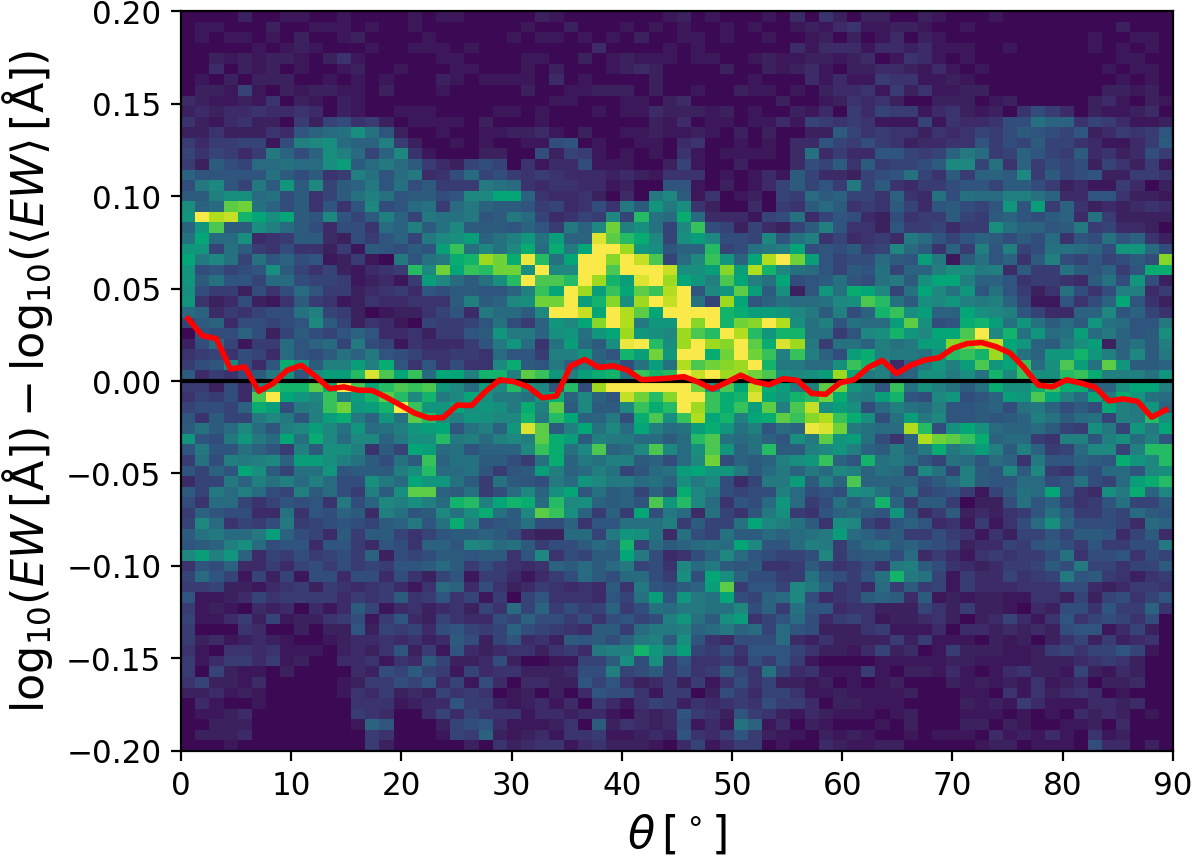}
\includegraphics[height = 0.25\textwidth]{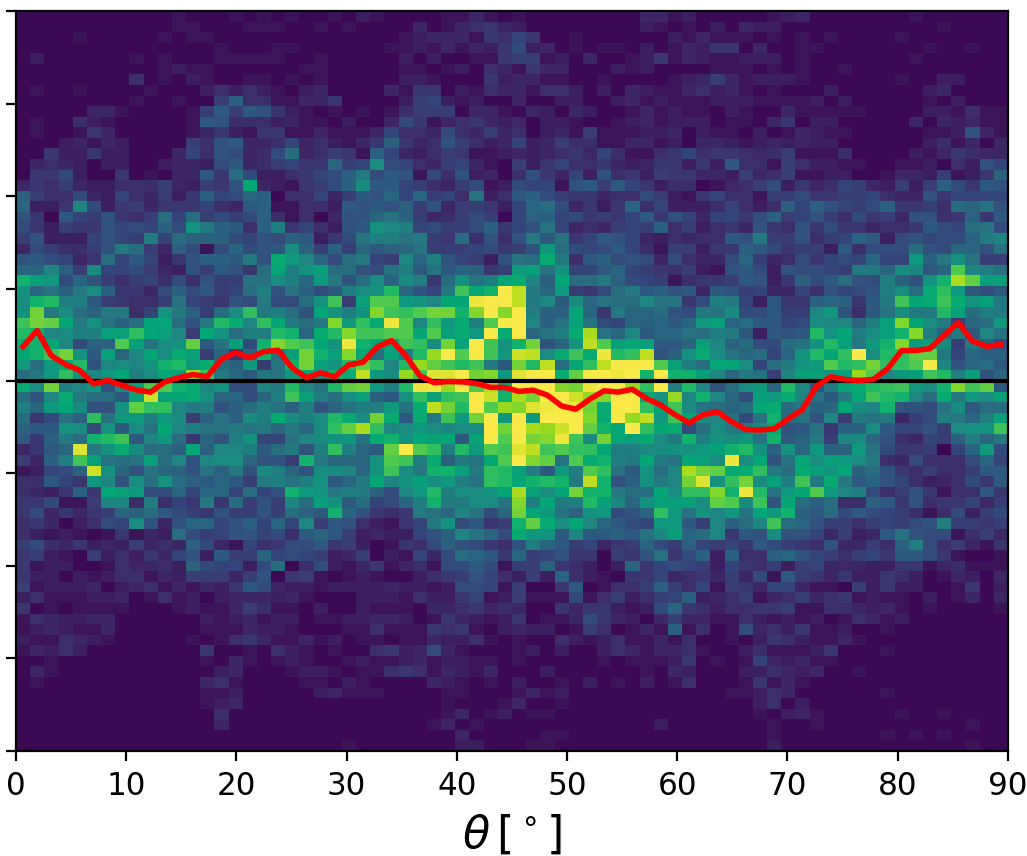}
\includegraphics[height = 0.25\textwidth]{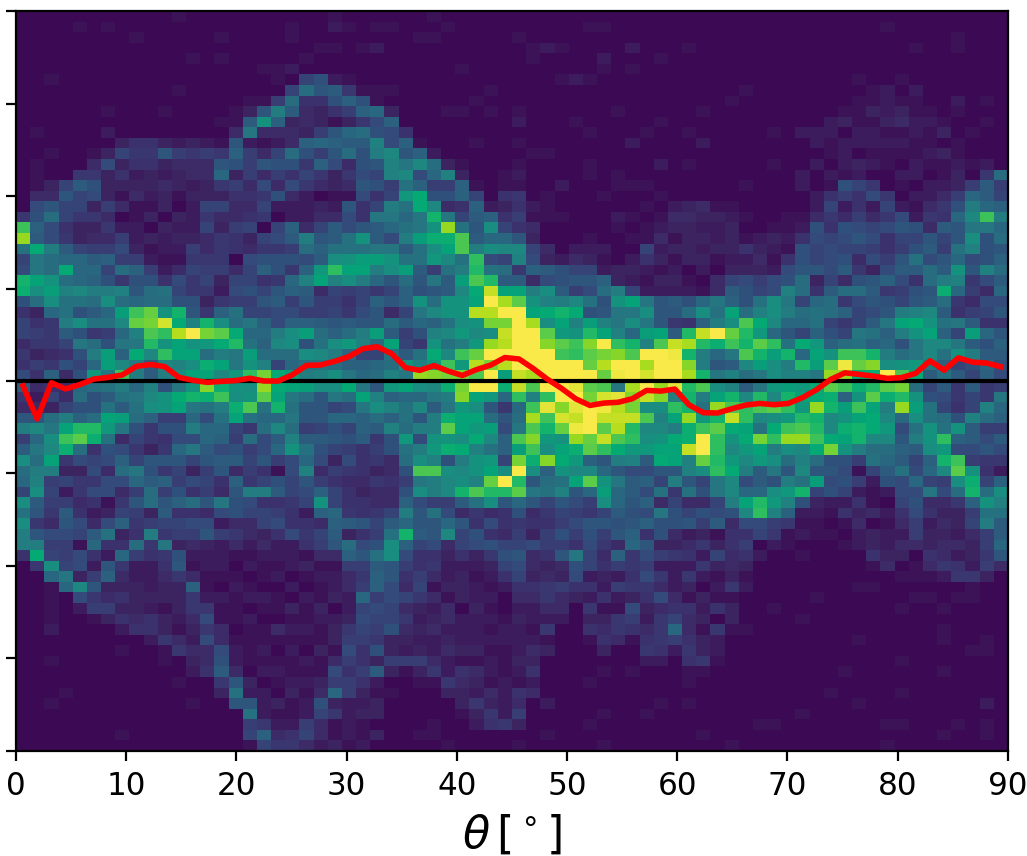}\\
\includegraphics[height = 0.25\textwidth]{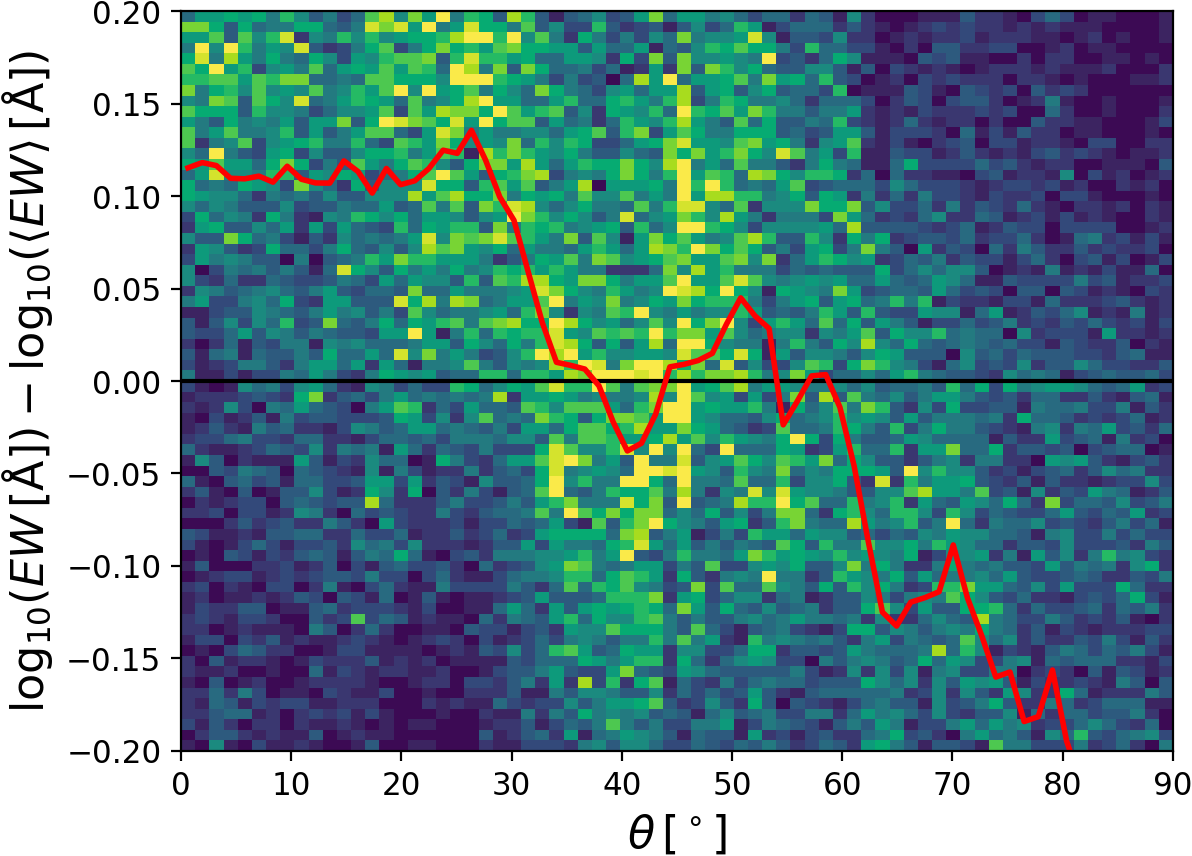}
\includegraphics[height = 0.25\textwidth]{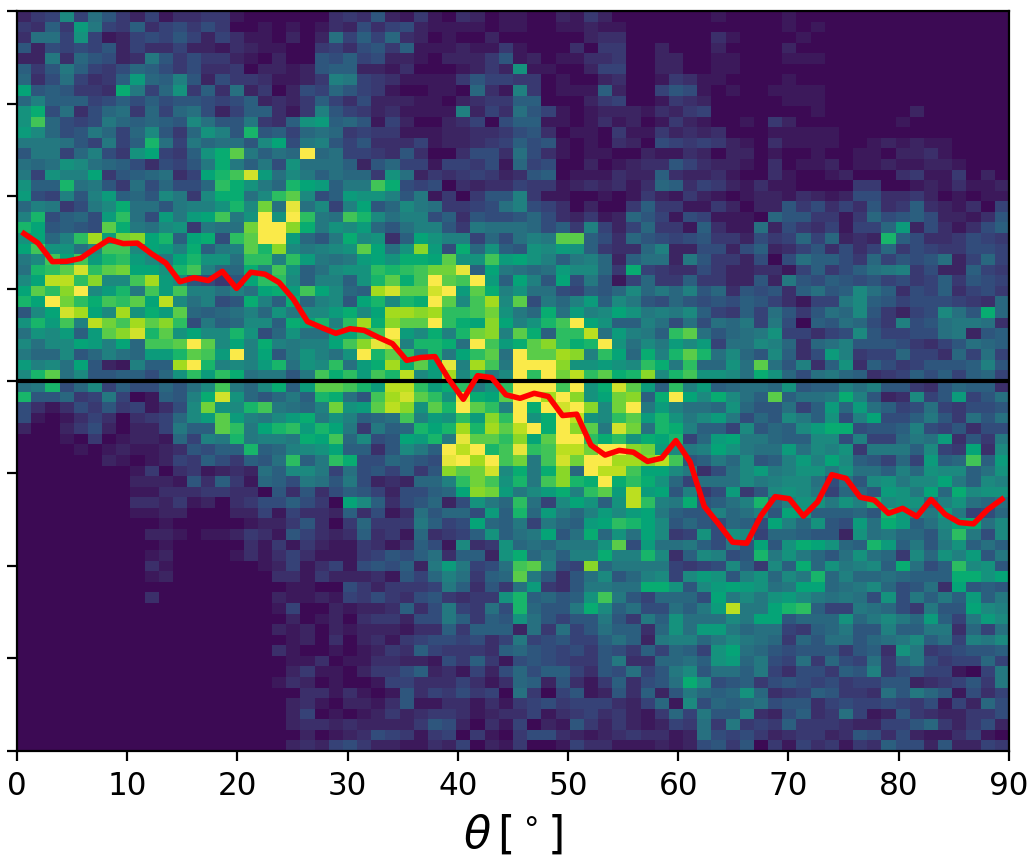}
\includegraphics[height = 0.25\textwidth]{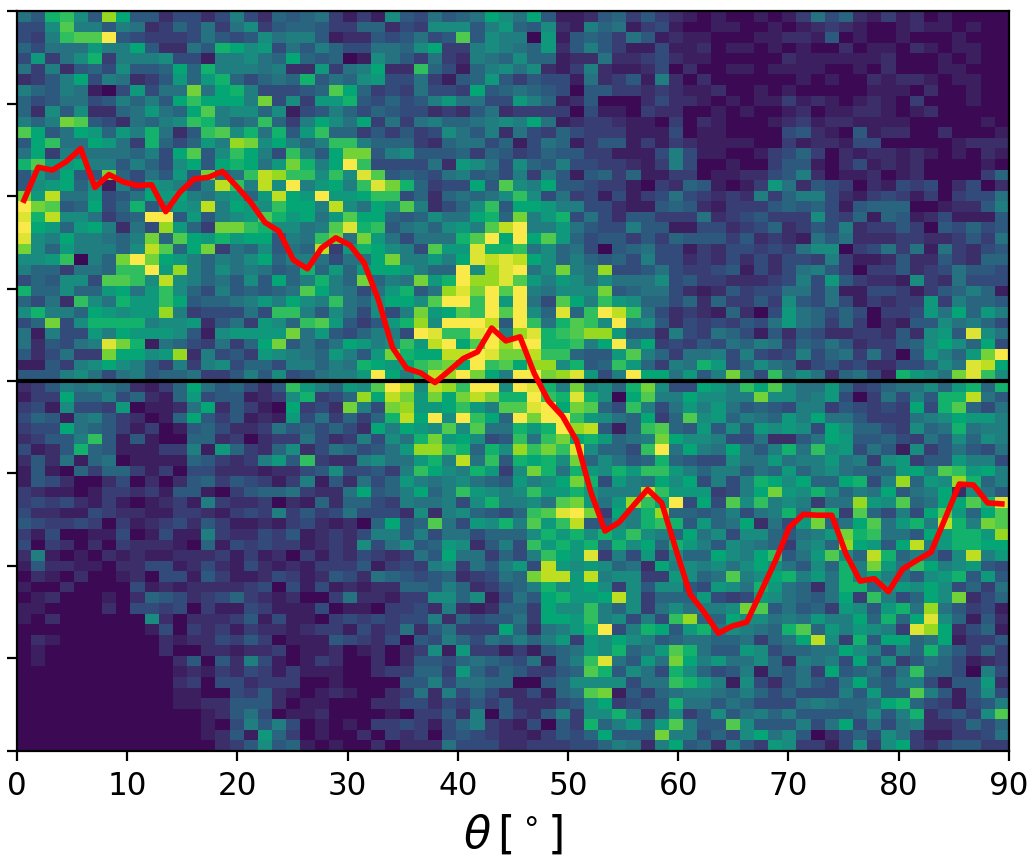}
\caption{
    Deviations of the EWs from the mean EW at a given radius as a function of the polar angle (the angle to the disc) of the LOS in edge-on projections. Here $0^\circ$ is in the plane of the disc and $90^\circ$ is perpendicular to the H\textsc{i} disc. The yellow colouring shows the density of LOS at that $\delta$EW and position angle.
	The red line is the median deviation at a given angle.
	We find two classes of halos: one without an angular dependence as is the case for M0664, M0858, and M2283 (upper row from left to right); and one with a clear angular dependence for which we show examples in the bottom row (from left to right: M0501, M0959, M1196).
}
\label{fig:EW_angular_dep}
\end{figure*}

\subsection{Angular Dependence of the Equivalent Widths}

We investigate the angular trends  of Lyman-$\alpha$ EWs which can be difficult to access with observations. The canonical view is that gas falling into the halo preferentially arrives in the disc plane, while outflows of generally hotter (enriched) gas are perpendicular to the disc. Hence higher EWs are expected in the disk plane relative to perpendicular to the disc.

Figure~\ref{fig:EW_angular_dep} shows the deviation of the EW from the radial average as a function of angle from the disc plane. We find two classes of objects. The first class are galaxies with clear signs of recent interactions or mergers show no correlation of EW variations with polar angle. These are simulations M0664, M0858, and M228, which are shown in the upper panels of Fig.~\ref{fig:EW_angular_dep}. Dense gas orbiting in the halo, also with increased velocity dispersion can result in elevated EWs at all polar angles at all radii. The typical effect of a minor merger or accretion event can be seen in M0858 (Fig.~\ref{fig:M0858_H1215_maps_zoom}). Note that a merger can disturb the system for a long time. For M1859 (Fig.~\ref{fig:M1859_maps_comp}) the last minor merger happened more than $5\,\mathrm{Gyr}$ ago. 

The second class of objects are galaxies in quiet environments like M0501, M0959, M1196 (bottom panels of 
Fig.~\ref{fig:EW_angular_dep}). They show clear correlations with polar angle, displaying high column densities extending to large radii well beyond the H\textsc{i} disc.  Hence in our simulations, galaxies must have experienced a very quiet merger history for a long time preceding observation in order to display a clear azimuthal signal of H\textsc{i} EW with position angle.

\section{Feedback and Resolution Dependence}
\label{sec:different_models}

To test the sensitivity of our results to variations in resolution and input physics, we run two additional simulations of Halo M1859. In the first, the `8x' run, we increase the mass resolution by a factor of 8 relative to our fiducial $4\times$ run.  We also run a weak feedback model (weakFB) where we reduce the outflow velocity in SNe from
$3000\,\mathrm{km/s}$ to just $30\,\mathrm{km/s}$, reducing the input of momentum
into the surrounding gas, which increasing the thermal energy input to maintain $10^{51}$~erg per SN.  We note that this run does not reproduce basic observations such as the stellar to halo mass ratio, but is still an instructive test case. Besides these changes, all physical modeling and analysis procedures are kept identical.

For the 8x run, the final halo properties are virtually identical to the 4x run: $M_{200}$ drops by 2.6\% and $R_{200}$ by $0.85\%$ relative to the 4x run. {For the stellar mass, however, the differences are larger. This}{The differences are larger when looking at the stellar and gas mass of the galaxy. Stellar mass} increases by $8.6\%$ up to $1.72\times10^{10}\,\mathrm{M_\Sun}$, whereas the gas mass decreases from 9.09 to $4.37\times10^9\,\mathrm{M_\Sun}$ ($-52.2\%$). {}{H\textsc{I} mass similarly decreases from 5.87 to $2.76\times10^9\,\mathrm{M_\Sun}$ (-53.1\%).}
The (mass-weighted) average stellar ages are again very similar: $6.96\,\mathrm{Gyr}$ for the 8x run versus $7.02\,\mathrm{Gyr}$ for the fiducial 4x run.  The drop in gas mass is much larger than the increase in stellar mass{, showing that the gas has ejected substantially more gas}{}.  Hence higher resolution tends to lock somewhat more mass into stars, but also blow out more gas.

The weakFB run shows much larger differences.  First, the halo mass grows by about 10\% (and correspondingly, $R_{200}$ by 3\%), demonstrating that feedback lowers the halo mass. The stellar mass increases by a factor of 4, to $6.74\times10^{10}\,\mathrm{M_\Sun}$, which shows that kinetic feedback from SN is crucial for regulating the conversion of gas into stars in the ISM.  Meanwhile, the gas mass drops to $2.98\times10^{9}\,\mathrm{M_\Sun}$, which is 3 times lower than the original run.  This directly arises because much of the gas has been converted into stars.  Hence strong feedback is responsible for keeping galaxies gas-rich.

\begin{figure*}
\centering
\includegraphics[width=\textwidth]{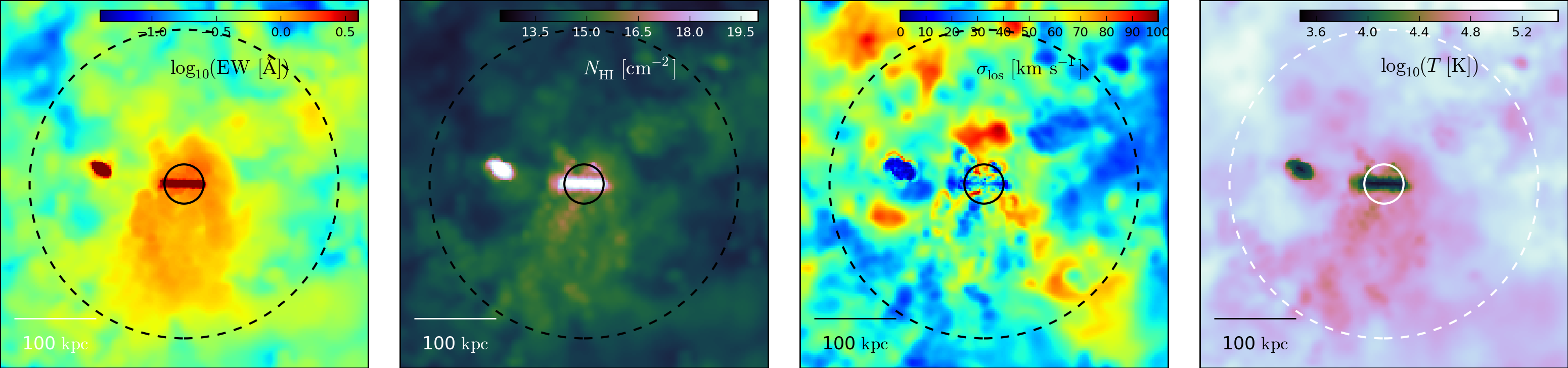}

\vspace{3mm}
\includegraphics[width=\textwidth]{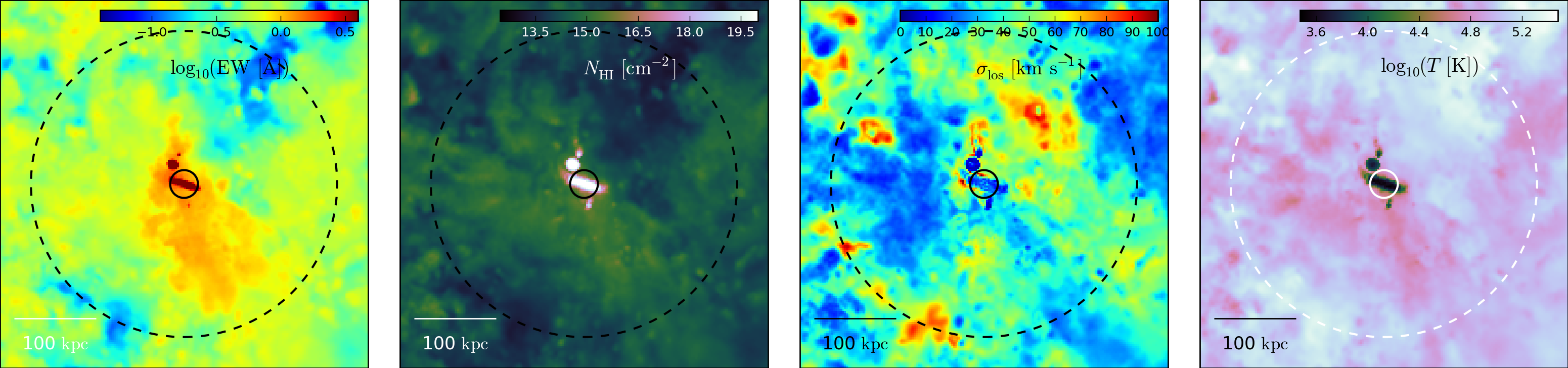}

\vspace{3mm}
\includegraphics[width=\textwidth]{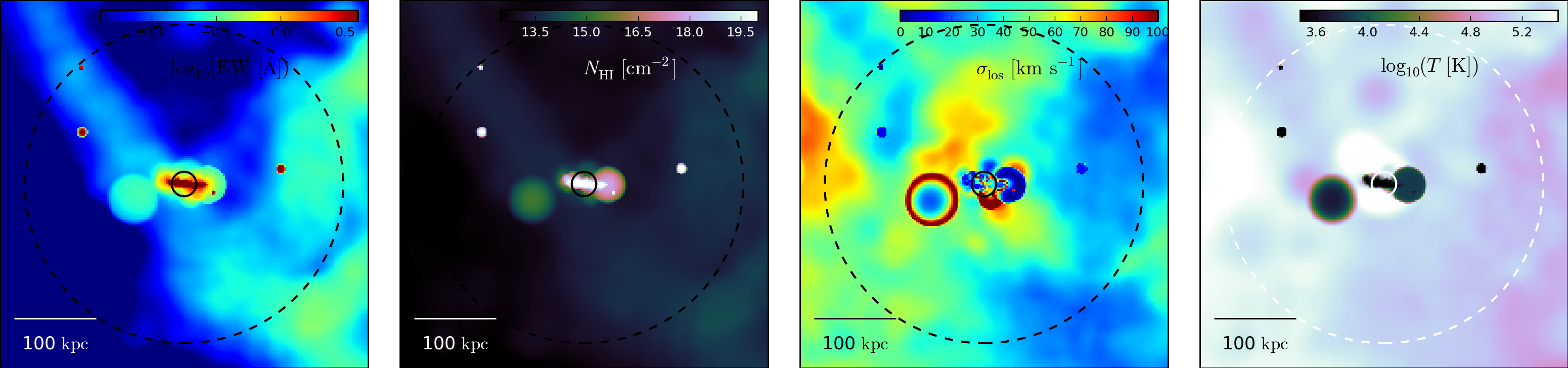}
\caption{
    The maps of equivalent width, hydrogen column density, line-of-sight dispersion, and temperature (from let to right) for M1859 at the standard resolution (top row) and at a factor 8 higher mass resolution (middle row). The simulation with weaker feedback at standard resolution is shown at the bottom row. Here the galaxy is almost devoid of a CGM as there is no gas recycling.}
\label{fig:M1859_maps_comp}
\end{figure*}

Figure~\ref{fig:M1859_maps_comp} shows maps for H\textsc{i} column densities, Lyman-$\alpha$ EWs, H\textsc{i} velocity dispersions, and temperature, for our three variants. The original 4x run is shown in the top row.   The 8x run in the middle row shows finer structures, but not overall a large difference in the general properties of the CGM.  However, the weakFB run shows dramatic differences, as the CGM is much emptier, to the point that single particle blobs of absorption become evident.  This shows that feedback has a dramatic impact on the structure of the CGM in these simulations.

%

\begin{figure*}
\centering            
\includegraphics[width=0.9\textwidth]{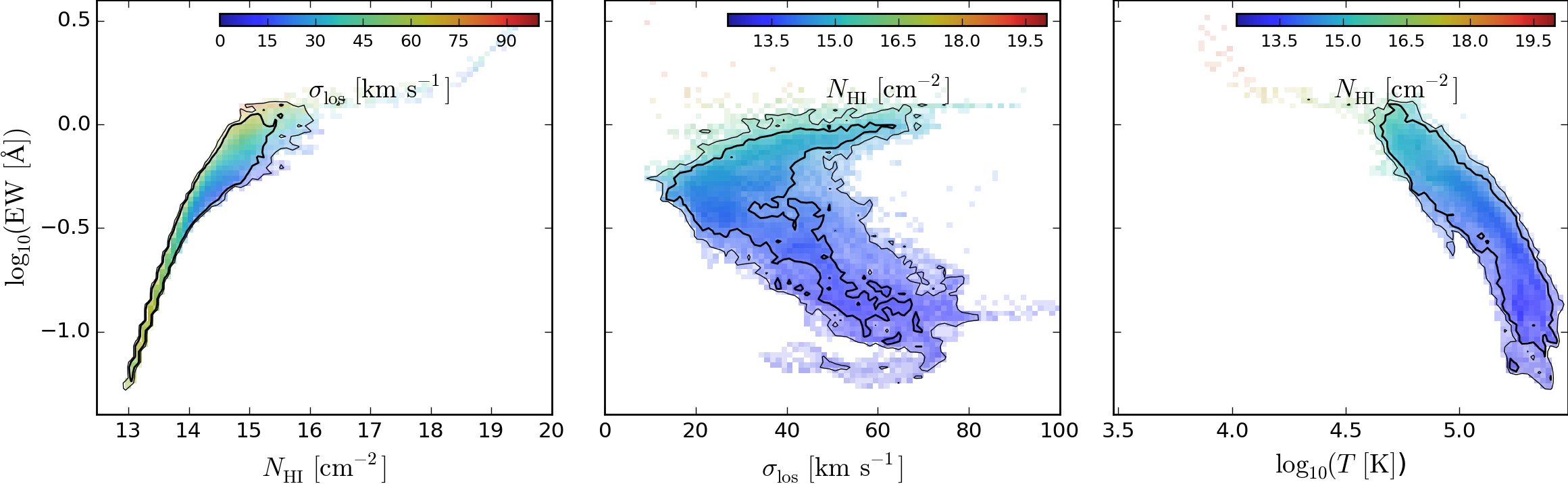}
\includegraphics[width=0.9\textwidth]{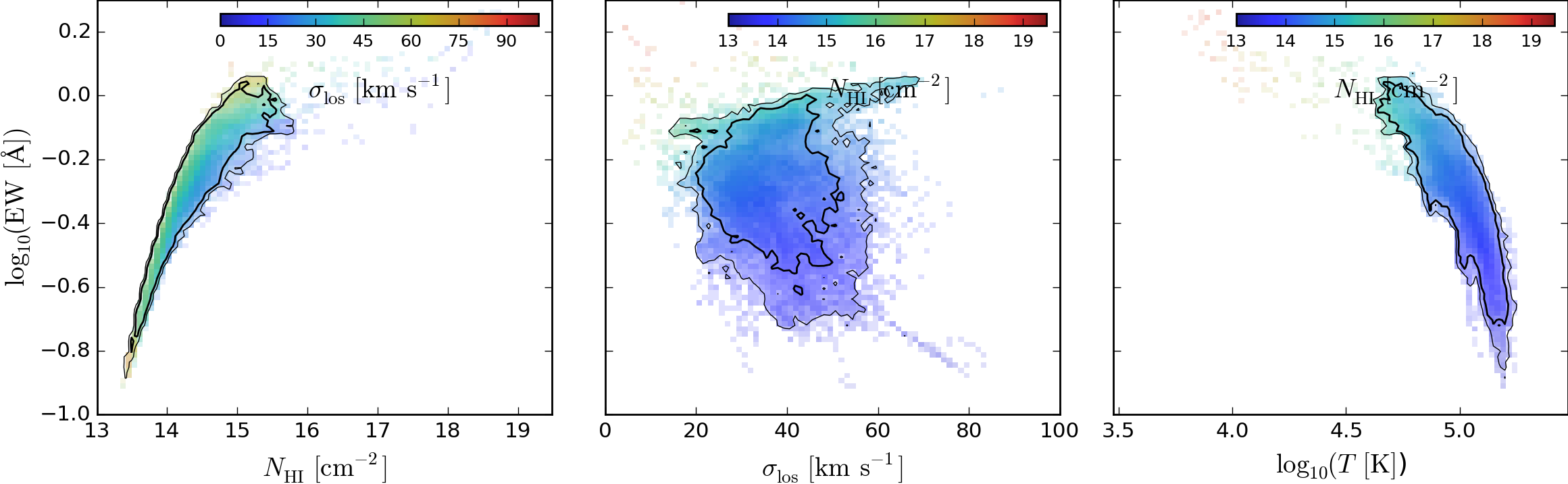}
\includegraphics[width=0.9\textwidth]{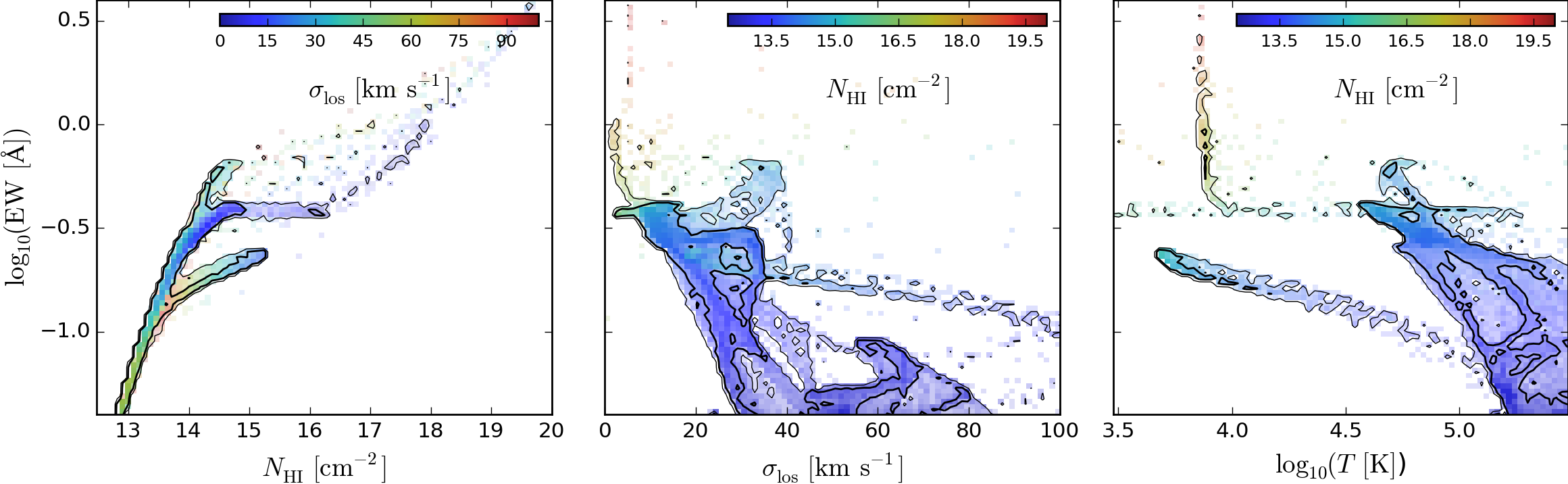}
\caption{
    The maps for M1859 with our standard model and resolution (upper left panel), with increased
    resolution (initial particle masses are larger by a factor of 8; upper right), and with the standard 
    resolution but a weaker feedback.
}
\label{fig:M1859_corr_comp}
\end{figure*}

Figure~\ref{fig:M1859_corr_comp} shows, the correlations of the EWs with column density, velocity dispersion, and temperature for all the discussed runs of halo M1859 that correspond to the maps in Fig.~\ref{fig:M1859_maps_comp}, similar to the plots shown in 
\ref{fig:M0858_H1215_corr_zoom}.

The first column shows a curve of growth plot.  Increasing resolution makes no strong change, but the weakFB run oddly shows three arms in the COG.  The lower arm corresponds to a single particle in the north-east corner of the maps in Fig.~\ref{fig:M1859_maps_comp}.  The upper two arms have EWs similar to the other two simulations and also their temperatures are
around $10^5\,\mathrm{K}$.

We also see differences in the line-of-sight velocity dispersion.  Increasing the resolution for this halo results in lower $\sigma_{los}$ in weaker systems, likely because these systems can now be resolved better.  The temperatures (right column) are also slightly lower, but this cannot explain the difference in the dispersions.  The weakFB run, in contrast, shows a dramatically different velocity and temperature structure, as much of the CGM has little to no absorption, and the strong lines are all concentrated to within the dense ISM.

\section{Summary}

We study the circumgalactic medium 11 zoom-in cosmological simulations of individual galaxy halos using an improved version of the SPH-based galaxy formation model of \cite{2013MNRAS.434.3142A}.  We exclude four that have mergers in their central galaxies, leaving 7 with with virial masses in the range of $M_{200}=4\times10^{11} - 3\times10^{12}\,\mathrm{M_\Sun}$, and central galaxy stellar masses of $7\times10^{9} - 1\times10^{11}\,\mathrm{M_\Sun}$, thus broadly consistent with the observed stellar to halo mass ratio; more properties are listed in Table~\ref{tab:haloprops}.  From these simulations we produced mock absorption line spectra for the  Lyman-$\alpha$ transition, using a new module in our publicly available python package \textsc{pygad}, following  \textsc{snapexbin}/\textsc{snapexsnap}~\citep{2006MNRAS.373.1265O,2010MNRAS.408.2051D}.  We study general characteristics of the CGM in these halos, and examine how Lyman-$\alpha$ absorption traces the physical and kinematic conditions of the CGM gas in these simulations.  Our main conclusions are as follows:
\begin{itemize}
    \item We find that our halos typically contain 50\% of their cosmic baryon fraction.  This is roughly divided equally between disc (ISM) gas, cool CGM gas, and warm-hot CGM gas.
    \item Halo accretion provides at least two-thirds of the CGM mass, and mass expelled from the ISM the remainder. This is qualitatively consistent with \citet{2019MNRAS.488.1248H}. At the present day, the rate of injection from these two sources is comparable.
    \item Most CGM gas that is cold ($T<10^{5.2}\,$K) was accreted onto the halo, and the majority of that in cold form.  ISM injected gas retained in the CGM was mostly injected hot.
    \item Accreted CGM gas dominates at radii larger than 50 kpc, while recyceled gas, gas expelled from the host galaxy ISM, dominates at smaller radii. The majority of CGM gas is accreted and not recycled within the host halo. 
    \item We find that the exact procedure for generating absorption lines from SPH particles can significantly impacts the results, particularly in saturated CGM absorbers, with differences of up to 20\% in EW. 
    \item Bulk motions provide a significant contribution to the EW of typical CGM absorbers, because they tend to lie in the logarithmic portion of the curve of growth.
    \item The strongest absorbers ($N_{HI}\ga 10^{18}\, $cm$^{-2}$) tend to arise in $10^4\, $K gas with low dispersion.
    \item The mean EW drops with increasing impact parameter, in rough agreement with observations though with a hint of too much absorption at large radii.
    \item Galaxies that experienced mergers show no azimuthal trend in EW, while galaxies that have had no mergers for at least $5\,$Gyr show significantly more absorption along the disc major axis and less along the minor axis.
    \item By comparing the EW and column density profiles, we find that CGM Lyman-$\alpha$ absorbers are best represented by a radially constant $b$ value that increases with halo mass, from $50\to70\,$km/s across our halo mass range. These values are larger than typically assumed but are consistent with \citet{2016ApJ...833..259B}.
    \item Our results are somewhat sensitive to resolution, with significantly less ISM gas in a higher resolution test case.  Our results are highly sensitive to feedback, as a run without feedback shows a dramatically different CGM that is strongly discrepant with observations.
\end{itemize}

In order to observationally disentangle the three main factors that contribute to the width of a typically saturated CGM Lyman-$\alpha$ line (column density, H\textsc{i} temperature, and LOS velocity) additional information beyond the Lyman-$\alpha$ line and its shape is needed.
One possibility is to look into the metal lines, which are  usually not saturated.
\cite{2015ApJ...802...10C} found that low-ionisation lines as such from Mg\textsc{ii} $\lambda2796$ are needed for this purpose as higher ionisation lines such as C\textsc{iv} $\lambda1548$ and O\textsc{vi} $\lambda1031$ originate from very distinct phases of the CGM.
By combining information from multiple ions together with insights from simulations such as these, it will be possible to disentangle the various CGM phases, obtain a more complete census of halo baryons, and trace the baryon cycle directly via absorption line probes.

\section{Acknowledgements}

This research was supported by the German Federal Ministry of Education and Research (BMBF) within the German-South-African collaboration project 01DG15006 "Ein kosmologische Modell f\"ur die Entwicklung der Gasverteilung in Galaxien". TN acknowledges support from the Deutsche Forschungsgemeinschaft (DFG, German Research Foundation) under Germany's Excellence Strategy - EXC-2094 - 390783311 from the DFG Cluster of Excellence "ORIGINS".
We thank Neal Katz for the helpful discussions and the \textsc{specexsnap/specexbin} code which we took as a basis for the implementation into \textsc{pygad}.

\section{Software used}
This work was made using NumPy \citep{doi:10.1109/MCSE.2011.37}, SciPy \citep{2020SciPy-NMeth}, Matplotlib \citep{Hunter:2007}, IPython \citep{PER-GRA:2007}, Jupyter notebooks \citep{Kluyver:2016aa}.

\bibliographystyle{mnras}
\bibliography{Lyman-alpha}

\label{lastpage}
\end{document}